\documentclass[fleqn,usenatbib]{mnras}

\usepackage{newtxtext}

\usepackage[T1]{fontenc}

\usepackage[usenames,dvipsnames]{xcolor}

\DeclareRobustCommand{\VAN}[3]{#2}
\let\VANthebibliography\thebibliography
\def\thebibliography{\DeclareRobustCommand{\VAN}[3]{##3}\VANthebibliography}

\usepackage[normalem]{ulem}
\usepackage{subfig}
\usepackage{graphicx}	
\usepackage{amsmath}	
\usepackage{amssymb}	
\usepackage{url}
\usepackage{balance}

\usepackage[most]{tcolorbox}
\usepackage{tikz,lipsum,lmodern}

\usepackage{mathtools}
\usepackage{siunitx}

\usepackage{fourier} 
\usepackage{array}
\usepackage{makecell}

\usepackage{orcidlink}

\usepackage{hyperref}
\usepackage{amsmath}
\usepackage[makeroom]{cancel}

\usepackage{algorithm,algpseudocode}
\usepackage{float}

\hypersetup{
    colorlinks = true,
    citecolor = {MidnightBlue},
    linkcolor = {BrickRed},
    urlcolor = {BrickRed}
}

\DeclarePairedDelimiterXPP\BigOSI[2]%
  {\mathcal{O}}{(}{)}{}%
  {\SI{#1}{#2}}
\DeclareMathAlphabet{\mathcal}{OMS}{cmsy}{m}{n}
\SetMathAlphabet{\mathcal}{bold}{OMS}{cmsy}{b}{n}

\title[Bayesian component separation for 21 cm IM]{Bayesian component separation and power spectrum estimation for 21 cm intensity mapping data cubes}

\author[Geoff G. Murphy et al.]{
Geoff G. Murphy,$^{1}$\thanks{E-mail: 4178310@myuwc.ac.za}\,\orcidlink{0000-0002-8186-3064}
Philip Bull,$^{2,1}$\,\orcidlink{0000-0001-5668-3101}
Mario G. Santos,$^{4,1}$\,\orcidlink{0000-0003-3892-3073}
Zheng Zhang,$^{2}$\,\orcidlink{0000-0002-9154-2803}
Steven Cunnington$^{3}$\,\orcidlink{0000-0001-6594-107X}
\\
$^{1}$Department of Physics and Astronomy, University of Western Cape, Cape Town 7535, South Africa\\
$^{2}$Jodrell Bank Centre for Astrophysics, University of Manchester, Manchester, M13 9PL, United Kingdom\\
$^{3}$Institute of Cosmology and Gravitation, University of Portsmouth, Portsmouth, PO1 2UP, United Kingdom\\
$^{4}$ South African Radio Astronomy Observatory, Black River Park, 2 Fir Street, Observatory, Cape Town, 7925, South Africa\\
}

\date{Accepted XXX. Received YYY; in original form ZZZ}

\pubyear{\the\year{}}

\begin{document}
\label{firstpage}
\pagerange{\pageref{firstpage}--\pageref{lastpage}}
\maketitle

\begin{abstract}
Foreground removal remains an ongoing challenge in radio cosmology, and increasingly sensitive experiments necessitate more robust analysis techniques. In this work, we model simulated data from a single-dish intensity mapping experiment, and use the Gibbs sampling and Gaussian constrained realisation (GCR) techniques to draw samples from the posterior probability distribution of the model parameters. This allows for a separation of the foregrounds and 21\:cm signal at the map level, as well as recovery of the 1-dimensional \textsc{Hi} power spectrum to within statistical uncertainties. Despite the model consisting of over 2\:million free parameters in the example presented here, these methods allow us to sample from the Bayesian posterior at a rate of $<30$ seconds per iteration. This framework is also resilient to frequency channel flagging (e.g. due to RFI excision), with the GCR steps effectively in-painting the missing data with statistically-consistent model realisations. The power spectrum is recovered accurately in the presence of strong foreground contamination and RFI flagging -- the estimate falling within $2\sigma$ of the true model in our example, similar to the commonly-used transfer function correction method. Statistical realisations of foreground and \textsc{Hi} maps are also recovered, with associated uncertainties available from the full joint posterior distribution of all parameters. 
\end{abstract}




\section{Introduction}\label{section:Introduction}
The redshifted 21\:cm line emission from neutral hydrogen (\textsc{Hi}) is a promising observational target for the study of cosmic large-scale structure. In principle, the ubiquity of \textsc{Hi} allows cosmological models to be tested across most of cosmic history, from the Dark Ages to the present day. With ever more sensitive instrumentation, increasingly faint emission can be detected, both at larger angular scales, and at higher redshifts. Combining sky coordinates with the redshift information encoded in the frequency dimension, the mapping of the \textsc{Hi} signal provides a three-dimensional tracer of matter, ionisation, and temperature \citep{Liu_Shaw_2020}, yielding information on the processes which produced the first stars and galaxies, as well as those that formed galaxy clusters, voids, and the cosmic web.

A central challenge for 21\:cm intensity mapping experiments is foreground contamination. Foreground contributions from Galactic synchrotron emission, free-free emission, and point sources such as distant galaxies can be orders of magnitude brighter than the \textsc{Hi} signal \citep{Mertens_2020, Hera_collab_2023}. The relative strength of each of these foreground sources is not known a priori, complicating the removal of the total foreground contribution, although efforts have been made to disentangle the individual contributions at microwave \citep{Thorne_2017,Zonca_2021} and radio wavelengths \citep{deOliveira-Costa_2008,Zheng_2017,Dowell_2017}.

Some experiments, such as the Hydrogen Epoch of Reionization Array (HERA), are designed to primarily observe the regions of Fourier space that are not contaminated by foregrounds, a method known as foreground avoidance \citep{Kerrigan_2018,Liu_Shaw_2020}. Instrumental and systematic effects can spread foreground power into modes that should be dominated by the 21\:cm signal however \citep{Aguirre+22,Murphy_2024, Rath_2024}, necessitating the modelling of foregrounds to some degree in order to undo the leakage. 

Many techniques have been proposed to separate the foregrounds from the \textsc{Hi} signal, but `blind' or data-driven foreground cleaning approaches are most commonly used by single-dish intensity mapping experiments. These use the structure of the data themselves, rather than explicit models, to separate out the foregrounds. Principal Component Analysis (PCA), for example, uses an eigenvector decomposition of the frequency-frequency covariance matrix of simulated or observed data to define an effective foreground model that is a linear combination of the dominant eigenvectors \citep{Alonso_2015}. Related methods include Independent Component Analysis (ICA) and Generalised Morphological Component Analysis (GMCA); see \citet{Spinelli_2022}.

The fact that foregrounds are spectrally smooth while \textsc{Hi} emission is rapidly spectrally varying should allow for fairly simple separation, but there does exist some overlap between these components, particularly on large radial scales. Some of the signal is therefore absorbed by the foreground eigenmodes, leading to signal loss -- the inadvertent over-subtraction of the 21\:cm signal when foreground cleaning is performed \citep{Burba_2024}. This must be corrected to avoid biased estimates of the \textsc{Hi} power spectrum, e.g. by calculating a transfer function describing the fraction of power lost as a function of scale \citep{Cunnington_2023}. This is worsened by instrumental effects that introduce additional spectral structure into the foregrounds, for example 
polarisation leakage \citep{Liao_2016, Harper_2018} and beam effects. This leads to the need for increasingly accurate (and complex) foreground {\it and} instrumental models, or methods capable of removing non-smooth foreground structure, e.g. GMCA \citep{Bobin_2007, Carucci_2020}, and mPCA \citep{Carucci_2025}.

In this work, we instead take a forward-modelling approach to the foreground separation problem. Explicit models of the foreground and signal components are formed, and compared with the data using a statistically rigorous Bayesian parameter estimation approach. This permits specific prior knowledge of instrumental effects such as beam patterns to be incorporated in the inference for example. The models are necessarily complex however; in what follows, we will study a simulated dataset that requires over 2 million free parameters to model it, including foreground eigenvector coefficients, \textsc{Hi} Fourier mode amplitudes, and their corresponding covariances (or power spectrum bandpowers).

Aside from defining a suitably accurate model, the problem then becomes to find a way of tractably estimating the parameter values despite the extremely large dimensionality of the parameter space. This is simply not possible using most statistical sampling methods due to the so-called \textit{curse of dimensionality}, a rapid decrease in performance of a sampling algorithm as more parameters are added \citep{Hu_2023,Peng_2024}.

Fortunately, this limitation can be avoided using specialised high-dimensional sampling methods. A well-known option is Hamiltonian Monte Carlo (HMC), which has previously been used to handle problems with (e.g.) hundreds of parameters in the context of 21\:cm cosmology \citep{Murphy_2024}. This is still several orders of magnitude away from the problem size of interest here however. Instead, we employ the combined methods of Gibbs sampling and Gaussian Constrained Realisations (GCR), which are capable of handling millions of free parameters if particular modelling choices are made -- principally, that large multivariate Gaussian sub-spaces of the posterior distribution exist, which can be accessed through a set of conditional probability distributions. As we will show, this allows us to draw samples (a single realisation) for all the parameters in $<30\:\rm s$ using one 32-core, 2.6 GHz Intel Xeon CPU node and less than $\sim4$\:GB RAM, which is thoroughly tractable with modern computing resources. These methods have been used extensively for modelling of cosmic microwave background maps \citep[e.g.][]{Eriksen_2008,Keihanen_2023}, and have also been applied to the reconstruction of full-sky foreground maps in interferometric radio observations \citep{Glasscock_2024}, and the joint calibration and formation of intensity maps \citep{2026RASTI...5ag024Z}, as well as in related applications to interferometric 21\:cm data \citep{2023ApJ...947...16A, Kennedy_2023, Burba_2024}.

The primary methodological comparisons for this work are the commonly-used data-driven foreground removal methods such as PCA, combined with a transfer function correction for signal loss \citep{2013ApJ...763L..20M, 2015ApJ...815...51S, Cunnington_2023, 2025MNRAS.542L...1C}. We discuss the transfer function estimation method in detail in Section~\ref{section:transfer_function}. In comparison, the framework we present in this work estimates the joint posterior distribution for the foregrounds, 21\:cm signal field, and its power spectrum directly, yielding a self-consistent model of the data that enables uncertainty propagation at the field level. This removes the need for a signal loss correction step; instead of the correlation (partial degeneracy) between foregrounds and spectrally smooth modes of the 21\:cm manifesting as a bias in the recovered power spectrum, it appears as covariance between the components in the recovered posterior distribution.

We test our framework on simulated autocorrelation data cubes, with the intention of applying it to the MeerKAT Large Area Synoptic Survey (MeerKLASS) in future. MeerKLASS is a \textsc{Hi} intensity mapping survey of the southern radio sky using the MeerKAT telescope in its single-dish configuration. The wide survey area and low resolution (but fast survey speed) makes it well-suited for cosmological surveys of unresolved 21\:cm emission. The MeerKLASS survey specifically aims to make a direct detection of baryon acoustic oscillations, survey continuum galaxies, and study cosmic large scale structure \citep{Santos_2017}. Due to substantial overlap in survey footprints, there is also the opportunity for cross-correlations with a number of other experiments, such as the Dark Energy Spectroscopic Instrument \citep[DESI;][]{DESI_2024}. MeerKLASS has already produced maps in MeerKAT's L-band; a pilot survey \citep{Wang_2021} allowed for a detection of the cross-power spectrum when foreground-cleaned intensity maps were correlated with the WiggleZ galaxy survey \citep{Drinkwater_2010, Cunnington_2022}. A cross-correlation detection using the Galaxy and Mass Assembly (GAMA) survey has also been obtained \citep{Meerklass_2025}. See \citet{Cunnington:2025sdr} for a recent review of MeerKLASS science results so far. Current observations are now being taken in the lower frequency (and relatively less RFI-contaminated) UHF-band of MeerKAT. 

Following this proof-of-concept implementation, our goal is to extend this method to include per-antenna systematics of the kind that are frequently encountered in single-dish experiments. In its current form, we are able to model the combined foreground, 21\:cm signal, and noise contributions to data cubes that have been calibrated, flagged for RFI, and averaged together from one or more antennas. This makes it directly applicable to existing high-level MeerKLASS data products for example. An important feature is the efficiency of the method in drawing samples from an extremely high-dimensional parameter space. Methods such as this are becoming ever more important as more sensitive and complex instruments such as the Square Kilometre Array (SKA) come online, with their correspondingly large data volumes.

This paper is structured as follows: Section\:\ref{section:Simulated_Data} describes the simulated data we use for our tests. Section\:\ref{section:models} outlines how we model the individual components, for example the foreground and \textsc{Hi} models. Section\:\ref{section:Bayesian_Model} describes our Bayesian sampling method, specifically how we draw samples for each component using the Gibbs sampling and GCR methods. Section\:\ref{section:Results} presents our results, which mainly consist of recovered data cubes and spherically averaged power spectra. Finally, we conclude in Section\:\ref{section:Conclusions}, as well as providing an indication of future work.

\section{Simulated Data}\label{section:Simulated_Data}
In this section, we detail the simulated data used to validate the method presented later in the paper. Foregrounds and 21\:cm cubes are both the same as the ones presented in \citet{Cunnington_2021}, whose simulations have a sky coverage of around $3\:000\rm \:deg^2$ (assuming a flat-sky approximation). The simulated data cubes have side lengths of $2048$ pixels, which are then rescaled to side lengths of $256$ pixels, with corresponding physical distances $\rm L_x,L_y,L_z=1000,1000,925\: \it h^{-1}\rm Mpc$. Here, $z$ is the radial (frequency) axis, with $x$ and $y$ being the transverse (angular) axes. The simulated maps are smoothed with a constant Gaussian beam with full-width at half-maximum (FWHM) corresponding to a MeerKAT-like dish diameter of $D_{\rm dish}=13.5\rm\: m$, referenced to a minimum frequency of $\nu_{\rm min} = 899\: \rm MHz$. The beam resolution is then
\begin{equation}
    \theta_{\rm FWHM} = \frac{1.18}{D_{\rm dish}} \frac{c}{\nu_{\rm min}} =1.67^\circ,
\end{equation}
with $c$ being the speed of light, and the prefactor being informed by the measured beam pattern of MeerKAT \citep{Matshawule_2021}. This beam treatment matches what was done in \citet{Cunnington_2021} to facilitate comparison; more realistic beams would have a FWHM that varies with frequency, and sidelobes with more complex spectral behaviour, however.

The \textsc{Hi} signal cubes from \citet{Cunnington_2021} are constructed from the \textsc{MULTIDARK-Galaxies} catalogue \citep{Knebe_2018}, derived in turn from the \textsc{MultiDark-Planck} $N$-body simulation \citep{Klypin_2016}. The latter simulated $3\:840^3$ dark-matter particles in a $1\:000^3 h^{-3}\rm\: Mpc^3$ cube. These simulations are trimmed along the line of sight to the aforementioned $1000,1000,925\: \it h^{-1}\rm Mpc$ to account for the assumed redshift range of $0.2<z<0.58$ (i.e. centred on $z=0.39$). The \textsc{MULTIDARK} data have coordinates in physical distances, and on a Cartesian grid, which \citet{Cunnington_2021} also adopt. 

The assumed cosmology for this simulation is $h=0.678$, $\rm \Omega_M=0.307$, $\rm \Omega_b=0.048$, $\rm \Omega_\Lambda=0.693$, $\sigma_8=0.823$, and an effective redshift of $z=0.39$. We opt to take only the first $128$ pixels along each dimension in order to reduce the problem size for our sampling test runs, while nevertheless retaining a realistic amount of data. This $128^3$ data, then, corresponds to a sky coverage of 730 $\rm deg^2$, with side lengths of $\rm L_x,L_y,L_z=500,500,415\:h^{-1}Mpc$. This removes the very largest modes where sky contamination is worst. In turn, this would hinder recovery of the power spectrum at the largest scales, but our focus here is on sub-matter-radiation equality scales.

For comparison, the MeerKAT data cube used in a recent cross-correlation detection of the 21\:cm signal had 199 frequency channels and approximately $30 \times 100$ angular pixels \citep{2023MNRAS.518.6262C}. For the majority of the results, we draw $2\:000$ samples, which takes approximately eight hours on a system with 32 CPU cores and $\sim4$ GB of RAM (where our code peaks at $
\sim2.5\:\rm GB$). Modelling of larger data cubes is tractable, but we do not perform computational scaling tests here.


The foregrounds of \citet{Cunnington_2021} are constructed from multiple sources, and include synchrotron emission, free-free emission, and point sources. They are similar to what would be observed in MeerKAT's L-band. A second foreground cube contains all of these components, plus a polarisation leakage effect that models Stokes Q and U-polarised synchrotron emission leaking into Stokes I as a result of magnetic field-induced Faraday rotation. In this work, we use only the leakage-free foregrounds; the inclusion of more complex effects, including chromatic beams, is left to future work.



We generate our own thermal noise cubes, assuming white noise (no correlated $1/f$ contribution) with zero mean that is constant over the observed volume. Different noise levels are used to test our method in different signal-to-noise regimes. We discuss this more explicitly in Section \ref{section:ps_21cm_recovery}, where we present the power spectrum recovery.

The simulated foregrounds and \textsc{Hi} define our `true' astronomical signal. Further adding a particular noise realisation gives a simulated data cube, which we denote by $\textbf{d}$.

\section{Data model}\label{section:models}

This section describes our approach to constructing and modelling the different components that we want to constrain: the amplitudes of a set of foreground modes;  3D Fourier mode amplitudes of the 21\:cm field; and the 1D 21\:cm power spectrum (i.e. the 21\:cm signal covariance). We describe the process of drawing samples from these components in Section \ref{section:Bayesian_Model}.

\subsection{Foregrounds}\label{section:foregrounds}

Our foreground model consists of a set of linear basis functions in frequency, with coefficients that are free parameters in each pixel. Any suitable linear basis may be used, for instance particular classes of polynomials, chosen up to some order that permits accurate reconstruction of the relatively foreground-dominated spectra.

\begin{figure}
    \centering
    \includegraphics[width=0.96\columnwidth]{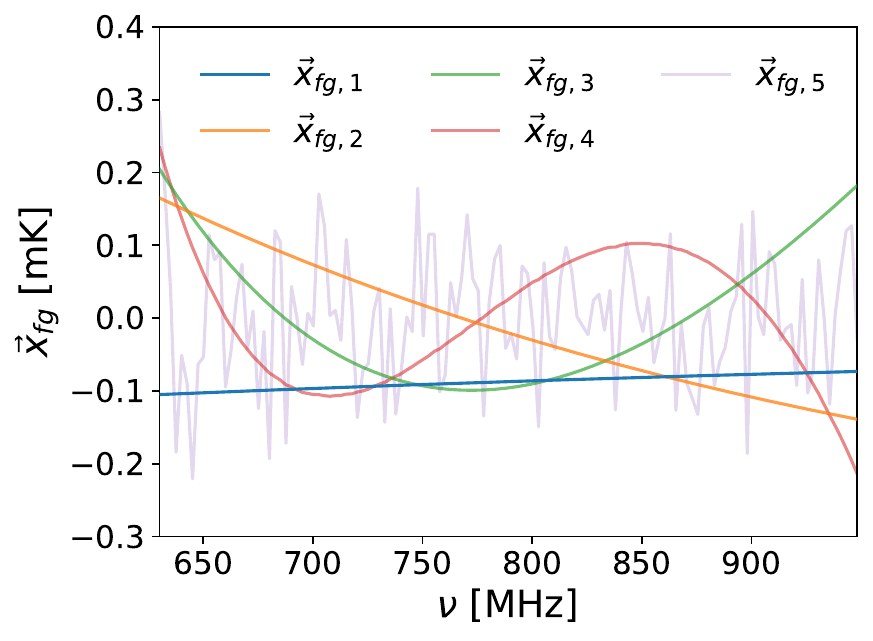}
    \caption{The first five foreground eigenvectors, of which only the first four are used in the model. }
    \label{fig:fg_evecs}
\end{figure}

For our tests here, we demonstrate a more data-driven choice of basis by adopting the leading-order PCA modes derived from the true (simulated) foreground maps. These are obtained by forming the frequency-frequency covariance matrix, averaged over all unmasked pixels, and then performing an eigendecomposition. Fig.~\ref{fig:fg_evecs} shows the first five eigenvectors,\footnote{For clarity, note that we use the terms ``eigenvectors'', ``PCA modes'', and ``foreground modes'' interchangeably throughout this paper.} 
of which only the first four are used. Higher order modes (i.e. the fifth onwards) are noise-like, suggesting they do not contain realistic foreground structure. Given that only single dish data are being modelled, the beam model is simple, and polarisation leakage is not included, the number of spectrally smooth modes is relatively small -- there is not a great deal of spectral structure to be modelled, so only a few modes are needed. Of course, more modes could be retained if more complex foregrounds are encountered.

The foreground model in each pixel is then a sum of these four eigenvectors, each a function of frequency, $\nu$, and each multiplied with a corresponding coefficient $f$, to yield a foreground model
\begin{equation}\label{eq:T_FG_ij}
    T_{{\rm FG}, ij} = \sum_{n=1}^{N_{\rm modes}} \mathit{f}_{n,i}\, {X}_n(\nu_j),
\end{equation}
where $i$ labels angular pixels, $j$ labels frequency channels, and $X_n$ is the basis function at order $n$.
For the data cube dimensions of $128^3$ voxels used here, we will therefore have $128^2$ 1D model foreground spectra, each with dimensions of $128$ frequency channels, resulting in $N_{\rm modes} \times N_{\rm pix} = 4 \times 128^2$ free parameters.

This is a serviceable if simplistic foreground model. It is practical to construct from simulations or even real data, and maintains the important property of being represented using linear basis functions, which is necessary for our sampling methods to work efficiently. It also involves a large number of parameters, which we will show can be tractably sampled. More sophisticated models can be readily adopted however, e.g. including some spatial information in the linear basis functions. The Galactic foregrounds in particular have non-stationary statistical properties on the sky (the Galactic plane looks very different to foregrounds at high Galactic latitude), and so sets of basis functions that are independent of angle are likely to give a less accurate reconstruction of the true foregrounds depending on direction. For present datasets, which cover relatively small regions of the sky and mostly avoid the Galactic plane, this is less important. A thorough exploration of different choices of foreground basis is left to future work. 


\subsection{Foreground mode-mode covariance}

Our statistical model permits prior distributions to be defined over the model parameters. For the foregrounds, this can be done in a few ways.

A useful approach is to define an angular power spectrum for the per-pixel foreground coefficients. Without this, the coefficients in each pixel are effectively fitted to the data independently, and are more likely to absorb some of the 21\:cm signal on large radial scales (i.e. small radial Fourier modes, $k_\parallel$). The foregrounds are known to have significant correlations on large angular scales however, and so it is reasonable to use a prior that enforces this.

Defining a Gaussian prior with a mode-mode and pixel-pixel covariance matrix of the form $F_{nn^\prime ii^\prime}$ is a possibility, although this will have a dense structure and is likely to be unwieldy to work with numerically. Instead, one can work with (Fourier or spherical harmonic) angular power spectra for each mode, $C_{\ell, n}$, and include transformations between the pixel and harmonic spaces to implement the prior in the sampling formalism developed below. We do not implement this approach in our proof-of-concept code, but it is mathematically straightforward to include, and numerically very similar to how we implement the power spectrum prior for the 21\:cm field (see below). 

In this work, we instead adopt a simpler prior that captures only the frequency-frequency structure of the foregrounds, which is the dominant correlation direction distinguishing them from the 21\:cm signal. We build the foreground covariance directly from the PCA decomposition of the foreground model itself, and take only the diagonal elements. This provides only limited prior information that restricts the maximum amplitude of each mode, and does not help to distinguish the foreground modes from the 21\:cm signal otherwise. 

A non-white foreground covariance (consisting of off-diagonal elements) assumes some level of stationarity in the sky, i.e. that the correlation structure is the same everywhere, which is not realistically the case for Galactic foregrounds (see Section \ref{section:foregrounds}). Enforcing priors on the off-diagonals could hinder recovery in parts of the sky where the assumed correlations are incorrect, so a diagonal prior is the more conservative choice.


\subsection{21 cm signal}\label{section:21cm_signal}


We model the 21\:cm field in 3D Fourier space, as this makes it more straightforward to relate to its power spectrum. The 21\:cm field itself is a real-valued brightness temperature field, which in Fourier space results in Hermitian (conjugate symmetric) complex coefficients. We denote these Fourier-space coefficients by $\mathbf{\tilde{{s}}}$, and use the real Fourier transform (i.e. the {\tt rfft} function) to implement the transforms between the coefficients and the real field. This requires only the coefficients with positive wavenumber, so we have $65\times128^2$ complex free parameters. We furthermore separate these into real and imaginary parts, so our free parameters are purely real, which is advantageous for use with some numerical libraries \citep{Glasscock_2024}. This results in a total of $2\times65\times128^2=2,129,920 $ free parameters for the 21\:cm signal in our model. 

\subsection{Signal power spectrum and covariance}

By modelling the 21\:cm field in terms of complex Fourier modes, and assuming statistical isotropy, the mode-mode signal covariance can be represented by a diagonal matrix with the binned power spectrum along the diagonal. We define $P(k_m)$ to be the bandpower centred on 1D wavenumber $k_m$. The 3D Fourier modes $\mathbf{k}$ that fall within a given 1D wavenumber bin share the same value of $P(k_m)$. Formation of power spectra from our actual data is discussed in Section \ref{section:Bayesian_Model}.


Estimating a covariance matrix normally requires many realisations of the underlying random field. As in any real observation, we have access to only a single realisation of the 21\:cm signal, so each Fourier mode provides just one sample of its own variance — too little information to estimate the diagonal of $\mathbf{S}$ mode-by-mode. We resolve this by binning in $k$. All Fourier modes falling within a given spherical shell are treated as independent realisations drawn from a common bandpower, $P(k_m)$, so that the many modes in each bin jointly constrain a single covariance entry.

The resulting covariance matrix $\mathbf{S}$ is diagonal in Fourier space, with one entry per complex-valued 21\:cm mode ($128^3$ in our model). Its entries take only $N_{k\:bins}$ distinct values — the binned bandpowers — so many modes share the same diagonal element. We choose $N_{k\:bins}$ = 14. These diagonal entries are bandpowers, so they are real and non-negative, and there is no need to separate real and imaginary components of $\mathbf{S}$.

Our full model, then, has $2,195,472$ free parameters, including the signal and foreground mode amplitudes, the signal power spectrum bandpowers, and the free foreground mode covariance elements.

\subsection{Noise covariance}

We generate our own noise cubes using particular standard deviations, $\sigma$, or equivalently noise root-mean-squares (RMS), $\sigma_{\rm RMS}$. We use the power spectrum signal-to-noise ratio (SNR) to define the noise levels -- this SNR is defined as the ratio of the signal bandpower to the noise bandpower in the highest $k$ bin, with the noise being a mean zero Gaussian random field, i.e.
\begin{equation}
    {\rm SNR}|_{k\approx0.45\:{\rm Mpc}^{-1}} = \frac{P(k)_{\rm HI}}{P( k)_{\rm Noise}}. \label{eq:noisesnr}
\end{equation}

Since the instrumental noise in a given experiment should be well-known a priori, the noise variance can be found with the square of the noise RMS, i.e. the covariance matrix
\begin{equation}\label{noise_N}
    \mathrm{N}_{q q^\prime} = \delta_{q q^\prime}\sigma_{\rm{RMS}, q}^2 = \delta_{q q^\prime}\sigma_{\rm{RMS}}^2,
\end{equation}
where $q$ labels individual voxels, and we have assumed that the noise variance is constant in frequency and pixel in what follows.

Furthermore, since the noise characteristics of the instrument are assumed to be known, $\mathbf{N}$ is constant and not sampled for. We do not consider other sources of noise, for example $1/f$ noise, but Gibbs sampling and GCR have been applied to this type of noise in \citet{2026RASTI...5ag024Z}. Lastly, while we assume constant $\sigma$, in observations noise does in fact change with frequency and sky position. The formalism presented in the next section is generalised to non-constant variance, but we did not include this in our simulations.

\section{Gibbs sampling formalism}\label{section:Bayesian_Model}
This section describes our priors, the likelihood function, and the process of drawing samples for our model parameters using Gibbs sampling and Gaussian Constrained Realisations. We also discuss our approach to frequency channel flagging, as well as our point of comparison for power spectrum recovery --- the transfer function correction.

\subsection{Likelihood and posterior distribution}

Our forward model is a function of the free parameters $\mathbf{p}$, and the model itself is denoted by $\mathbf{m(p)}$. To compare the model with the target data, $\mathbf{d}$, one evaluates a likelihood function. We choose a Gaussian likelihood, along with an assumption of Gaussian random noise, i.e.
\begin{equation}\label{eq:likelihood}
    \rm{log}\,\mathit{p} (\mathbf{d}\,|\,\mathbf{p}) \propto -\frac{1}{2}(\mathbf{d - m(p)})^T\mathbf{N}_w^{-1}(\mathbf{d-m(p)}).
\end{equation}
We define the inverse noise covariance as $\mathbf{N}_{\rm w}^{-1} \equiv \mathbf{w^T N^{-1}w}$, where the weights, $\mathbf{w}$, take into account missing or downweighted data. Data excision in the process of RFI removal is one example where the weighting term is used. Flagged pixels have a corresponding value of $0$ in $\mathbf{w}$, while unflagged (retained) pixels have a value of $1$.

Bayes' theorem states that the posterior distribution is equal to the product of the likelihood and prior, normalised by the evidence \citep{Trotta_2017},
\begin{equation}
    p(\mathbf{p\,|\,d)} = \frac{p(\mathbf{d\,|\,p)\mathit{p}(\mathbf{p)}}}{p\mathbf{(d)}} , \label{eq:BayesProp}
\end{equation}
where the evidence is typically ignored under a proportionality sign.

We can write our model for the data cube as
\begin{equation}
    \mathbf{m = U_s\, s + U_f\, f},
\end{equation}
where $\mathbf{U_s}$ is the operator which projects the Fourier amplitude vector $\mathbf{s}$ into a model for the real \textsc{Hi} field. This is simply \textsc{numpy}'s {\tt rfft} function as described in Section \ref{section:21cm_signal}. Similarly, $\mathbf{U_f}$ projects $\mathbf{f}$, the foreground mode amplitude vector, into a model of the foreground cube — i.e. it applies the foreground basis functions $X_n$ to the mode amplitudes, as given explicitly in Eq. \ref{eq:T_FG_ij}. The projection operators can be explicitly represented as matrices, but it is often more efficient in the numerical implementation to use a `matrix-free' approach, e.g. using the much faster FFT functions that also take up less memory.

We have intentionally constructed our signal and foreground model components using linear bases, so that the model is linear in their amplitude parameters. This leads to the amplitude parameters having a (multivariate) Gaussian conditional distribution under this choice of Gaussian likelihood function, as we will see in the next sub-section. This is the key modelling assumption that permits efficient high-dimensional sampling of the corresponding sub-space of the parameter space. To preserve this Gaussian structure of the conditional distributions, we must also choose either uniform or Gaussian priors for these parameters. In the latter case, they can be written as
\begin{align}\label{eq:priors}
\begin{split}
    p(\mathbf{s,f)} &= p(\mathbf{s)}p(\mathbf{f}) \\
    &\propto \exp\left({-\frac{1}{2}(\mathbf{s-\overline{s}})^{\rm{T}}\mathbf{S^{-1}}(\mathbf{s-\overline{s}})}\right)/\mathbf{\,|\,S\,|\,^{\frac{1}{2}}} \\
    &\times \exp\left({-\frac{1}{2}(\mathbf{f-\overline{f}})^{\rm{T}}\mathbf{F^{-1}}(\mathbf{f-\overline{f}})}\right)/\mathbf{\,|\,F\,|\,^{\frac{1}{2}}},
\end{split}
\end{align}
where the signal and foreground priors are assumed independent. The prior means are denoted by $\overline{\mathbf{s}}$ and $\overline{\mathbf{f}}$, and their covariances by $\mathbf{S}$ and $\mathbf{F}$. In this work, the signal priors are chosen to be uninformative and mean-zero, $\overline{\mathbf{s}}=\mathbf{0}$, while the foreground priors, $\overline{\mathbf{f}}$, are informative --- produced by applying a 10\% random Gaussian offset to the true foreground amplitudes. We discuss these choices further in Sect.~\ref{section:priors}. We will see shortly that $\mathbf{S}$ and $\mathbf{F}$ can themselves be sampled from separate conditional distributions.

We can now write the joint posterior distribution of our model as
\begin{align}\label{eq:joint_pos}
\begin{split}
        p(\mathbf{s,f,S,F\,|\,d}) &\propto p(\mathbf{d\,|\,s,f,S,F})p(\mathbf{s,f,S,F}) \\
        &\propto p(\mathbf{d\,|\,s,f,S,F)}p(\mathbf{s})p(\mathbf{f})p(\mathbf{S})p(\mathbf{F}).
\end{split}
\end{align}
This distribution is not a multivariate Gaussian, and is intractable to sample from directly owing to the large number of parameters. 

\subsection{Gibbs sampling and Gaussian Constrained Realisations}



We now employ \textit{Gibbs sampling}, where samples are iteratively drawn from conditional distributions that are each more tractable than the joint posterior distribution itself. The conditional distributions are not necessarily Gaussian, but it is helpful when a large fraction of the parameters occupy (conditional) multivariate Gaussian sub-spaces, as samples can be drawn from these directly and efficiently even when the number of parameters is very large. In this paper, we call the method of drawing samples from these sub-spaces `Gaussian Constrained Realisations' (GCR). This does away with the need to use Markov Chain (MCMC) sampling methods for these conditional distributions, which are approximate, take time to converge, and also suffer from the curse of dimensionality. Sub-spaces with non-Gaussian conditional distributions and small numbers of parameters can still use MCMC methods if needed however. Other `direct' sampling methods also exist for a select few non-Gaussian distributions, such as the inverse Gamma distribution discussed below. 


The process of Gibbs sampling can be summarised as follows: instead of trying to jointly sample all of the parameters in Eq. \ref{eq:joint_pos} as
\begin{equation}
    \rm \mathbf{s}_{q+1},\mathbf{f}_{q+1},\mathbf{S}_{q+1},\mathbf{F}_{q+1} \leftarrow p(\mathbf{s}_q,\mathbf{f}_q,\mathbf{S}_q,\mathbf{F}_q\,|\,\mathbf{d}),
\end{equation}
where the full distribution itself is not Gaussian, we instead iteratively sample from the simpler conditional distributions as
\begin{align}\label{eq:GCR}
    \begin{split}
        \mathbf{s}_{q+1}, \mathbf{f}_{q+1} &\xleftarrow{} p(\mathbf{s,f}\,|\,\mathbf{S}_q,\mathbf{F}_q,\mathbf{d}) \\
        \mathbf{S}_{q+1} &\xleftarrow{} p(\mathbf{S}\,|\,\mathbf{s}_{q+1},\mathbf{d}) \\
        \mathbf{F}_{q+1} &\xleftarrow{} p(\mathbf{F}\,|\,\mathbf{f}_{q+1},\mathbf{d}) .
    \end{split}
\end{align}
At iteration $q$ of the sampling, the samples $\mathbf{s}_{q+1}$ and $\mathbf{f}_{q+1}$ are drawn from the conditional distribution $p(\mathbf{s,f}\,|\,\mathbf{S}_q,\mathbf{F}_q,\mathbf{d})$. Samples for the signal covariance modes, $\mathbf{S}_{q+1}$ are then drawn in the next step using the newly drawn $\mathbf{s}_{q+1}$ samples, marginalising over the parameters of the foregrounds and its covariance. This can be done because $\mathbf{S}$ is only dependent on the signal modes, $\mathbf{s}$, and not on $\mathbf{f}$ or $\mathbf{F}$. 

A similar step is then carried out to draw samples for the foreground covariance modes, $\mathbf{F}_{q+1}$, which is dependent only on $\mathbf{f}_{q+1}$. The loop then begins again with the updated parameter values, drawing the samples $\mathbf{s}_{q+2}$ and $\mathbf{f}_{q+2}$, using the newly drawn samples of the covariance modes, etc. Vitally, all three steps of this scheme involve known distributions that can be randomly sampled from exactly, rather than using MCMC methods that are subject to the curse of dimensionality.

\subsection{GCR step for 21 cm and foreground amplitudes}

Step 1 of Eq. \ref{eq:GCR} jointly samples for the amplitudes of the complex 21\:cm Fourier modes, and the foreground PCA modes. The prior spaces of both of these are Gaussian (by construction), and so the peaks of these distributions can be solved for directly.

Following \citet{Eriksen_2008}, we substitute our likelihood from Eq. \ref{eq:likelihood} and priors from Eq. \ref{eq:priors} into Bayes' theorem (Eq.~\ref{eq:joint_pos}), and then inspect the conditional distribution for the amplitude parameters, which is
\begin{align}\label{Eq:posterior}
    \begin{split}
        p(\mathbf{s,f\,|\,S,F,d}) &\propto p(\mathbf{d\,|\,s,f,S,F})p(\mathbf{{s},f\,|\,S,F}) \\
        &\propto \exp\left({-\frac{1}{2}(\mathbf{d-m(s,f)})^{\rm T} \mathbf{N_w^{-1}}(\mathbf{d-m(s,f)})}\right) \\
        &\times \exp\left({-\frac{1}{2}(\mathbf{{s}-\overline{s}})^{\rm{T}}\mathbf{S^{-1}}(\mathbf{s-\overline{s}})}\right) \\
        &\times \exp\left({-\frac{1}{2}(\mathbf{f-\overline{f}})^{\rm{T}}\mathbf{F^{-1}}(\mathbf{f-\overline{f}})}\right).
    \end{split}
\end{align}
The determinants of the covariance matrices, $\,|\,\mathbf{S}\,|\,$ and $\,|\,\mathbf{F}\,|\,$, are ignored under the proportionality sign as these are fixed in this conditional distribution, and so only act as scaling factors here.

Following \citet{Eriksen_2008}, Eq.~\ref{Eq:posterior} can be written in a simple multivariate Gaussian form with block vector and matrix terms,
\begin{equation}
     p(\mathbf{s,f\,|\,S,F,d}) \propto \mathrm{exp}\left( -\frac{1}{2}(\mathbf{x-\hat{x}})\mathbf{A}(\mathbf{x-\hat{x}}) \right).
 \end{equation}
Following \citet{Kennedy_2023}, one can derive a linear equation $\mathbf{Ax=b}$, where
\begin{equation}\label{eq:A}
\mathbf{A} = \begin{pmatrix}
    \mathbf{S^{-1} + U_s^T N_w^{-1} U_s} & \mathbf{U_s^T N_w^{-1}U_f} \\
\mathbf{U_f^T N_w^{-1} U_s} & \mathbf{F^{-1} + U_f^T N_w^{-1}U_f} 
\end{pmatrix},
\end{equation}
and 
\begin{equation}\label{eq:B}
    \mathbf{b} = \begin{pmatrix}
        \mathbf{U_s^T(N^{-1}wd + N^{-\frac{1}{2}}w^{\frac{1}{2}}}\omega_d) + \mathbf{S^{-1}\overline{s} + S^{-\frac{1}{2}} }\omega_s\\
        \mathbf{U_f^T(N^{-1}wd + N^{-\frac{1}{2}}w^{\frac{1}{2}}}\omega_d) + \mathbf{F^{-1}\overline{f} + F^{-\frac{1}{2}} }\omega_f 
    \end{pmatrix}.
\end{equation}
This can be solved to give a random realisation of the block vector of amplitude parameters, $\mathbf{x} = (\mathbf{s}, \mathbf{f})$.

To implement sampling in our model, we include the terms $\omega_s$, $\omega_f$, and $\omega_d$, which are unit Gaussian random vectors with the same shape as their corresponding components, $\mathbf{s}$, $\mathbf{f}$, and $\mathbf{d}$, respectively. Setting these $\omega$ terms to zero, the linear equation solution is the maximum (conditional) a posteriori estimate (MAP), i.e. the peak of the distribution. By implementing the sampling terms, the solutions fill in, or explore, the posterior space around this MAP solution, and produce the typical distributions one associates with posteriors, revealing degeneracies, correlations, etc.

For $\omega_s$, since $\mathbf{s}$ is complex, we include separate unit Gaussian random draws for the real and imaginary parts, but respecting the correct Hermitian symmetry. In contrast, $\omega_f$, and $\omega_d$ are both real.

\subsection{Preconditioner for the GCR step}\label{sec:precond}

The linear system defined by Eqs.~\ref{eq:A} and \ref{eq:B} can be quite large and numerically challenging to solve. To aid in convergence, we use a preconditioner --- see for example \citet{Saad_2003} and \citet{Eriksen_2004}. This is a constant matrix $\mathbf{M}^{-1}$, which multiplies both sides of the linear system so that the equation $\mathbf{M}^{-1}\mathbf{Ax}=\mathbf{M}^{-1} \mathbf{b}$ is solved instead. Good preconditioners can be computed rapidly, and will reduce the dynamic range of the eigenvalues of the linear operator, which permits convergence in fewer iterations. For instance, a preconditioner that gives $\mathbf{M}^{-1} \mathbf{A} \approx \mathbf{I}$ would greatly reduce the time needed to solve the system.

From Eq. \ref{eq:A}, we can derive a suitable preconditioner by taking only the diagonal elements
\begin{equation}
    \mathbf{M} = \begin{pmatrix}
    \mathbf{S^{-1} + N_w^{-1}} & 0 \\
0 & \mathbf{F^{-1} + U_f^T N_w^{-1}U_f}
\end{pmatrix},
\end{equation}
where $\mathbf{U_s^T N_w^{-1} U_s}$ reduces to $\mathbf{N_w^{-1}}$ because we choose ortho-normalised Fourier transforms\footnote{Implemented via \texttt{np.fft.rfftn(..., norm=`ortho')} in \textsc{numpy}.} for $\mathbf{U_s}$. We lose the off-diagonals as these are low-rank coupling terms, so $\mathbf{M}^{-1} \mathbf{A}\approx\mathbf{I} + \delta$ --- the identity matrix plus a low-rank perturbation, which is the type of system where Krylov methods like \textsc{lgmres} converge very quickly.

Since $\mathbf{S}$ and $\mathbf{N}$ are diagonal in Fourier space, we can invert the signal block on a per mode basis:
\begin{equation}
    \mathbf{M}_{00}^{-1}[k] = \frac{1}{1/\mathbf{S}_k + 1/\mathbf{N}_k}.
\end{equation}
For the foreground block, the uninverted expression is 
\begin{equation}
    \mathbf{M}_{11} = \text{diag}(1/\mathbf{F}_i) + \frac{1}{\mathbf{N}} \, \mathbf{x} \mathbf{x}^{\rm{T}},
\end{equation}
where $\mathbf{x}$ are the PCA eigenvectors or equivalent orthonormal foreground modes. We can find $\mathbf{M}_{11}$ via direct inversion because $n_{\rm modes}=4$, so this is not an expensive operation.

In the code, the preconditioner is defined as a linear operator and is simply passed to the solver, i.e. \texttt{lgmres(..., M=M\_inv)}.

\subsection{Sampling step for the 21 cm covariance}\label{section:GCR_S}

Following the GCR step for the amplitudes of the \textsc{Hi} field and foregrounds described above, the signal covariance matrix, $\mathbf{S}$, is sampled by conditioning on the newly drawn sample of $\mathbf{s}$. Since it is only dependent on this parameter vector, the conditional probability can be written as 
\begin{equation}\label{eq:S}
    p(\mathbf{S_{q+1}\,\,|\,\,s_{q+1}}) \propto\frac{1}{\,|\,\mathbf{S}\,|\,^{\frac{1}{2}}}\exp\left( -\frac{1}{2}(\mathbf{s}_{q+1}-\overline{\mathbf{s}})^T\mathbf{S^{-1}}(\mathbf{s}_{q+1}-\overline{\mathbf{s}}) \right).
\end{equation}
Samples of $\mathbf{S}$ cannot be drawn in general because it is statistically underdetermined for a single realisation of $\mathbf{s}$. Our bandpower parametrisation ensures that multiple 3D Fourier modes share the same bandpower parameters however, i.e. they belong in the same $\rm \,|\,\mathbf{k}\,|\,$ bin used to determine the bandpowers of the power spectrum $P(k)$. Fourier modes in the same bin are then considered to be realisations of the same underlying quantity, allowing for a statistically sound calculation of the signal covariance. $\mathbf{S}$, then, has $\rm{N_{\rm k\:bins}}$ free parameters --- the number of chosen power spectrum bins.

In Fourier space (denoted by tildes), the covariance is given by 
\begin{equation}
    \mathbf{\tilde{S}}\equiv \langle \mathbf{\tilde{s}^T\tilde{s}}\rangle,
\end{equation}
where angle brackets denote the ensemble average, which in this case is estimated from a mean over all Fourier modes in a particular bin. $\mathbf{\tilde{S}}$ is real and diagonal given that we model the \textsc{Hi} signal in Fourier space, so
\begin{equation}\label{eq:s=pk}
    \mathbf{\tilde{S}}_{jj^\prime} = \delta_{j j^\prime} P(k_j).
\end{equation}
For clarity, $\mathbf{s}$ is the vector of real-space \textsc{Hi} voxel amplitudes, $\mathbf{\tilde{s}}$ the Fourier space \textsc{Hi} voxel amplitudes, and $\mathbf{\tilde{S}}$ the covariance matrix formed from the latter. Note that we have only parametrised the 1D power spectrum here, although it is common to model the 21\:cm power spectrum as a function of both radial and transverse Fourier modes, $P(k_\perp, k_\parallel)$. An extension to this `cylindrical' power spectrum parametrisation is left to future work. 

It can be shown that Eq.~\ref{eq:S} has the form of a product of inverse Gamma distributions, one for each bandpower parameter represented in the diagonal matrix $\mathbf{\tilde{S}}$ \citep{Kennedy_2023}. The inverse Gamma distribution is defined as \citep{Gelman_2013}
\begin{equation}\label{eq:inv_gamma}
    \mathbf{f(x;\alpha,\beta)} = \frac{1}{\Gamma(\alpha)} \left(\frac{1}{x}\right)^{\alpha+1}\mathrm{exp}\left( -\frac{\beta}{x} \right),
\end{equation}
with $\alpha=(N_k/2)-1$, $x=\mathbf{S}$, and $\beta=\frac{1}{2}(\mathbf{s} -\mathbf{\overline{s}})^{\rm T}(\mathbf{s} -\mathbf{\overline{s}})$. $\Gamma$ is the Gamma function. Making the substitutions and using the diagonality of $\mathbf{\tilde{S}}$, we can then write the conditional probability as
\begin{align}\label{eq:S_prod}
\begin{split}
    p(\mathbf{S}\,|\,\mathbf{\tilde{s}}) &\propto \frac{1}{\,|\,\mathbf{\tilde{S}}\,|\,^\frac{1}{2}}\mathrm{exp} \left(-\frac{1}{2} \mathbf{\tilde{s}}^T \mathbf{\tilde{S}}^{-1} \mathbf{\tilde{s}}\right) \\
    &=\prod_{k}[P_k]^{-\frac{N_k}{2}}\exp\left(-\frac{1}{2}\frac{\sigma_k^2}{P_k} \right),
\end{split}
\end{align}
where $N_k$ denotes the number of Fourier modes that contribute to a bandpower $P_k$ with index $k$, and where we have defined
\begin{equation}
    \sigma_k^2 \equiv \sum_{\,|\,\mathbf{k}_j\,|\,\in \mathcal{B}_k}\mathbf{\tilde{s}}_j^*\mathbf{\tilde{s}}_j,
\end{equation}
which amounts to a (scaled) variance estimate. Here, $\mathcal{B}_k$ denotes the set of Fourier modes within the bin defined for bandpower $k$.

By drawing from each of the independent inverse Gamma distributions in the product, we can obtain samples of the bandpowers, $x=P_k$.

Note that the conjugate prior for an inverse Gamma distribution is another inverse Gamma distribution. It is common to implement this prior by modifying the shape parameter $\alpha$ \citep{Glasscock_2024}. For a uniform prior, the shape parameter is simply $\alpha=\frac{1}{2}N_{k}-1$, which is adopted throughout this work. Adjusting this prior (for example by setting $\alpha \to \alpha + 1$) can help with foreground-signal separation at large scales, and other issues related to biased bandpower estimates, but we did not find this to be the case in this work.

\subsection{Sampling step for the foreground covariance}\label{section:GCR_F}

As with the signal covariance, the foreground covariance is only dependent on $\mathbf{f}$, the foreground mode amplitudes. This again allows us to sample from the conditional distribution $p(\mathbf{F}\,|\,\mathbf{f})$, with no direct dependence on the 21\:cm signal parameters or the data vector.

For a single foreground mode, we define the vector $\mathbf{f}^{(i)}$, which groups all amplitude coefficients corresponding to that mode --- i.e. $i\in\{1,2,3,4\}$. For a statistically stationary foreground amplitude distribution, it can then be assumed that all $\mathbf{f}^{(i)}$ are realisations drawn from the same underlying distribution. 

Defining $\overline{\mathbf{f}}$ as the prior mean for a particular foreground mode amplitude, the conditional probability can then be written as
\begin{align}
\begin{split}
    p(\mathbf{F}\,|\,\mathbf{f}) &\propto \prod_i \frac{1}{\,|\,\mathbf{F}\,|\,^{\frac{1}{2}}} \exp \left(-\frac{1}{2}(\mathbf{f}^{(i)} - \overline{\mathbf{f}}^{(i)})^T \mathbf{F}^{-1} (\mathbf{f}^{(i)} - \overline{\mathbf{f}}^{(i)})\right)\\
    &= \prod_i \frac{1}{\,|\,\mathbf{F}\,|\,^{\frac{1}{2}}} \exp \left [-\frac{1}{2}\rm{Tr}(\mathbf{F}^{-1}\mathbf{D}^{(i)})\right]\\
    &=\,|\,\mathbf{F}\,|\,^{-\frac{N_{\rm pix}}{2}}\exp\left[ -\frac{1}{2}\mathrm{Tr}(\mathbf{F}^{-1}\widetilde{\mathbf{D}}) \right],
\end{split}
\end{align}
where
\begin{equation}
 \mathbf{D}^{(i)} \equiv (\mathbf{f}^{(i)} - \overline{\mathbf{f}}^{(i)})   (\mathbf{f}^{(i)} - \overline{\mathbf{f}}^{(i)})^T.
\end{equation}
The scale matrix $\widetilde{\mathbf{D}}$ is given by
\begin{equation}
    \widetilde{\mathbf{D}}\equiv \frac{1}{N_{\rm pix}}\sum_i^{N_{\rm pix}}\mathbf{D}^{(i)} ,
\end{equation}
with $N_{\rm pix}$ being the total number of pixels. $\mathbf{F}$ can then be sampled by drawing from an inverse-Wishart distribution (the multivariate form of the inverse-Gamma distribution),
\begin{equation}
    W^{-1}_p(\mathbf{X;}\Psi,\nu) = \frac{1}{\Gamma_p(\frac{\nu}{2})}\,|\,\mathbf{X}\,|\,^{-\frac{(\nu+p+1)}{2}}\exp\left(-\frac{1}{2}\mathrm{Tr}(\Psi\mathbf{X^{-1}}) \right),
\end{equation}
where $\Gamma$ is the gamma function, $p=\rm{n_{\rm modes}}$ (the number of foreground modes used in the model), $\nu=N_{\rm pix}-p-1$, $\mathbf{X}=\mathbf{F}$, and $\Psi=\widetilde{\mathbf{D}}$. Here, $\nu$ is the number of degrees of freedom, and $p$ is the dimension of the scale matrix, which equals the number of foreground modes. 

As an aside, if one were only concerned with the recovery of summary statistics, i.e. the foreground covariance and the power spectrum bandpowers, then the amplitude parameters would be considered nuisance parameters which could be analytically marginalised over (with closed-form integrals, as the distributions are Gaussian). We would then have some marginal likelihood described by $p(\mathbf{d}|\mathbf{S},\mathbf{F})$, reducing the number of parameters to $\sim14$ bandpowers plus the foreground covariance elements.

This likelihood would depend on a dense post-marginalisation covariance of shape $\rm N_{data}\times N_{data}$ however. This requires inversion at certain steps --- precisely what this work aims to avoid, especially for large dense matrices such as $\mathbf{S}$. Furthermore, with $\mathbf{S}$ now being contained in the post-marginalisation covariance matrix, the bandpowers would lose their conjugate inverse-Gamma form, preventing the direct sampling we employ here, and instead requiring MCMC. The trade-off would be far fewer parameters, but with a potentially more expensive sampling step. Whether this would result in overall faster sampling is not immediately apparent. The `full posterior evaluation' approach presented here also has additional benefits, such as map-level reconstruction, RFI in-painting, and the option of including effects which are not commonly evaluated with summary statistics, e.g. different types of systematic effect.

\subsection{Priors}\label{section:priors}

For the most part, our model priors are fairly simple. The Fourier mode prior mean is chosen to be $\overline{\mathbf{s}} = \mathbf{0}$, i.e. uninformative, leaving the signal to be recovered as a statistical fluctuation. The foreground amplitude priors are derived from the true amplitudes, multiplied by a 10\% Gaussian random offset: $\overline{\mathbf{f}} = \mathbf{f}_{\rm True}\times\mathcal{N}(\mu=1, \sigma=0.1)$. This assumes a realistic scenario in which there is some a priori knowledge of the foregrounds, to within 10\% of their true values. As discussed in Section \ref{section:GCR_S}, the signal covariance shape parameter is $\alpha=\frac{1}{2}N_{k}-1$, which is equivalent to a uniform prior, and similarly, from Sect.~\ref{section:GCR_F}, the foreground covariance has shape parameter $\nu = \rm N_{pix}-p-1$.

\subsection{Frequency channel flagging}

We also test the model on the same simulations, but with frequency channels pseudo-randomly removed. This emulates RFI excision. The frequency flags were defined manually, and were chosen to include both narrow and broad regions, as well as closely and sparsely-spaced regions. As set up here, this corresponds to $\sim28\%$ of the data being flagged.

The model maintains the same setup as for previous results, with only the weight vector $\mathbf{w}$, and weighted noise covariance, being updated to take into account the missing data (Eqs. \ref{eq:likelihood}, \ref{eq:A}, and \ref{eq:B}). This is a vector with the same number of pixels as the data, with corresponding values of one at unflagged pixels, and zero in flagged pixels. Apart from this, the model is run with the same settings as in the unflagged case.

\subsection{Transfer function}\label{section:transfer_function}

Our primary point of comparison for our \textsc{Hi} power spectrum recovery is the transfer function-corrected power spectrum of PCA foreground-subtracted data. This approach is frequently used in \textsc{HI} intensity mapping experiments, MeerKLASS included. The transfer function corrects for signal loss following blind foreground removal by injecting simulated/mock signal into the foreground-contaminated data. Thereafter, the mock plus data combination is again foreground-cleaned, and the resulting signal loss in the mock component provides an estimate of the loss in the true signal, which is then corrected for. The process can be summarised as below, with \cite{Cunnington_2023} providing a more detailed discussion:

\medskip
\noindent(i) Clean the foregrounds in the observed/true data via PCA. 

\medskip
\noindent(ii) Calculate the power spectrum of the cleaned data, $P_{\rm clean}(k)$.

\medskip
\noindent(iii) Create mock signal cubes (for which we use FastBox\footnote{\url{https://github.com/philbull/FastBox}}).

\medskip
\noindent(iv) Inject this mock signal into the foreground-contaminated data, and apply PCA-cleaning to this set of data. Thereafter, subtract the foreground-cleaned data-only component from step (i).

\medskip
\noindent(v) Calculate the transfer function with
\begin{equation}\label{eq:transfer_function}
    \mathcal{T}(k) = \left\langle \frac{\mathcal{P}(\mathbf{X}^{\rm{m}}_{\rm{clean}},\mathbf{X}_{\rm{m}})}{\mathcal{P}(\mathbf{X}_{\rm{m}},\mathbf{X}_{\rm{m}})}\right\rangle_{N_{\rm{mock}}} .
\end{equation}
\medskip
\noindent(vi) Correct for signal loss in the cleaned data by multiplying with the inverse of the transfer function,
\begin{equation}
    P_{\rm rec}(k) = P_{\rm clean}(k)[\mathcal{T}(k)]^{-1} .
\end{equation}
$\mathcal{P}(\mathbf{X}^{\rm{m}}_{\rm{clean}},\mathbf{X}_{\rm{m}})$ in Eq. \ref{eq:transfer_function} is the cross-correlation power spectrum between the cleaned mock + data component, $\mathbf{X}^{\rm{m}}_{\rm{clean}}$, and the mock signal, $\mathbf{X}_{\rm{m}}$. The denominator is the autocorrelation power spectrum of the mock data. An ensemble average is taken over $N_{\rm{mock}}$ realisations of the simulated \textsc{Hi} signal, where $N_{\rm{mock}}=1000$ for this work.

While very computationally efficient, a disadvantage of the transfer function correction is that the number of PCA modes required for good recovery is not well-defined. Some trial-and-error is required in order to find a number of PCA modes which does not lead to under or overcorrection when applying $\mathcal{T}(k)$ to the cleaned data. Something similar could be said of our approach, where the choice of modes/basis functions can lead to worse component separation, or induce degeneracies with the signal field. For the transfer function, however, this choice of basis is baked into both the PCA cleaning and the correction function, so some of the effects \citep[such as mode-mixing;][]{2025MNRAS.542L...1C} are partially absorbed into the correction. Our approach does provide some feedback on the choice of number of modes, particularly in terms of goodness of fit statistics and worsening degeneracies between signal and foreground modes as the foreground model becomes more flexible. There is also the possibility of doing model comparisons using Bayesian evidence ratios, which would provide a quantitative `best-choice' for the number of PCA modes, although the evidence is not a direct output of the Gibbs sampling approach and would have to be estimated through other means.

For this work, four PCA modes were used to foreground-clean the true data, and in the cleaning of the mock \textsc{Hi} + data components when forming the transfer function.

\section{Numerical implementation}
\label{sec:implementation}

This section summarises several important numerical implementation
choices, adapting the framework discussed in Section \ref{section:Bayesian_Model}.

\subsection{State vectors, operators, and constrained realisation solver}

At each iteration we draw the joint constrained realisation
$(\mathbf{s},\mathbf{f})$ by solving a single linear system
$\mathbf{A}\mathbf{x}=\mathbf{b}$. The state is implemented as one real
1D array,
\begin{equation}
  \mathbf{x} = \bigl[\,\mathrm{Re}(\tilde{\mathbf{s}}),\;
                      \mathrm{Im}(\tilde{\mathbf{s}}),\;
                      \mathbf{f}\,\bigr]^{\mathsf T},
\end{equation}
where $\tilde{\mathbf{s}}$ are the real-FFT signal modes and
$\mathbf{f}$ are the $\rm N_\mathrm{mode}=4$ PCA foreground amplitudes.
The signal and foreground projection operators $\mathbf{U}_s$ and
$\mathbf{U}_f$ are never formed explicitly; they are implemented as functions (\texttt{Us}, and \texttt{Uf}) that apply
\texttt{numpy.fft.rfftn}/\texttt{irfftn} (with \texttt{norm=`ortho'}),
and a PCA mode projection respectively, with a \texttt{transpose}
flag selecting the forward or adjoint action.

Similarly, the full
operator $\mathbf{A}$ is not explicitly formed, but is instead implemented as a function which evaluates the result of applying each of the 2$\times$2 blocks on the input block vector $\mathbf{x}$, as per Eq. \ref{eq:A}. The output is a 1D vector concatenating the results of this block operation. Implementing the block matrix-vector operation in this way (a so-called `matrix-free' approach) has major speed and memory efficiency advantages.


The right-hand side, $\mathbf{b}$, is also assembled using a function, taking the parameter means, the data, etc., as input. The output is a 1D vector, the concatenation of $({\mathbf{b}_0, \mathbf{b}_1 )}$. The linear system is wrapped in a \texttt{scipy.sparse.linalg.LinearOperator} and solved with
\texttt{lgmres} to a relative tolerance of $10^{-8}$, warm-started
from the previous iteration's sample. A block-diagonal
preconditioner $\mathbf{M}^{-1}$, built once per iteration, is supplied via the \texttt{M} argument, as discussed in Sect.~\ref{sec:precond}.

\begin{figure}
    \centering
    \includegraphics[width=0.85\columnwidth]{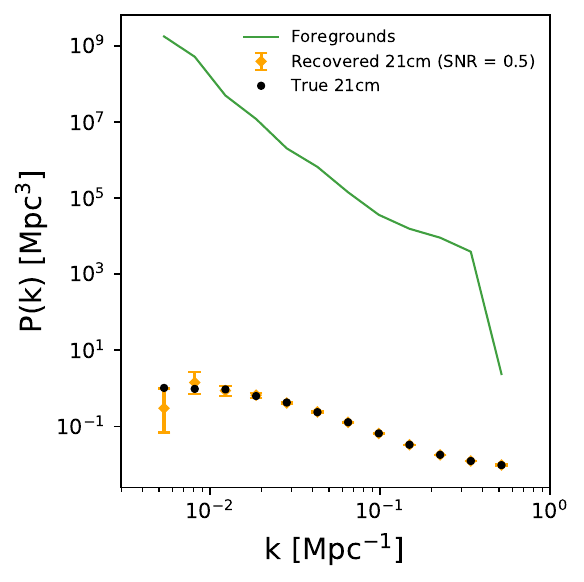}
    \caption{An example of the \textsc{Hi} power spectrum recovery in comparison to the foreground power. The black circles denote the true \textsc{Hi} power spectrum, the green line the foregrounds, and the orange diamonds our power spectrum samples with statistical uncertainties (95\% CL errorbars).}
    \label{fig:recovery_w_fg}
\end{figure}

\subsection{Covariance updates}

Given the newly drawn $(\mathbf{s},\mathbf{f})$, the two
covariance parameter blocks are updated independently. The signal power
spectrum $P(k)$ is sampled using the \texttt{scipy.stats.invgamma} function, applied separately to the 14
$k$-bins that are defined once at setup from the full 3D $|\mathbf{k}|$ grid.
The resulting $P(k)$ values are broadcast back onto every mode in
each bin to form the diagonal matrix $\mathbf{S}$, with the DC mode (which cannot be sampled using the inverse Gamma distribution) artificially set
to a large value ($10^{30}$) so that
$\mathbf{S}^{-1}[0]$ has no contribution.

\begin{figure*}
    \centering
    \includegraphics[width=1.75\columnwidth]{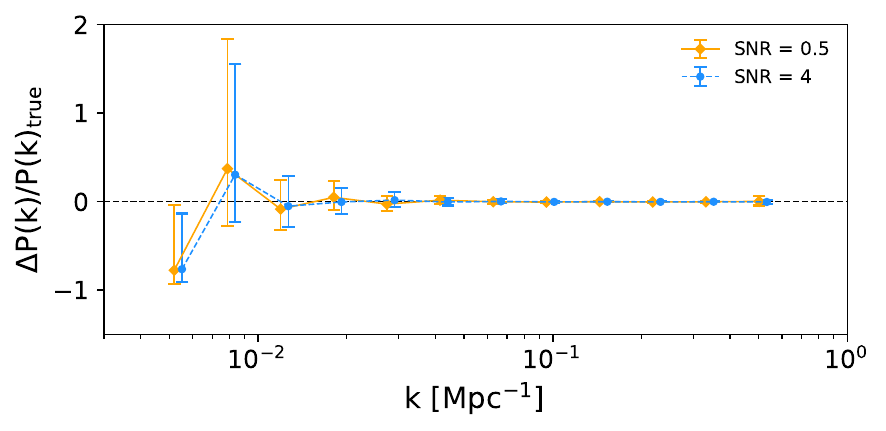}
    \caption{The recovery of the true power spectrum shown as the fractional residuals for the case of power spectra SNR values of 0.5 and 4 --- orange solid line and blue dashed line, respectively. These residuals are calculated with respect to the median power spectra, with their associated 95\% CL statistical uncertainties being derived from 1250 samples. Points have been offset slightly in $k$ for legibility.}
    \label{fig:snr_comp}
\end{figure*}

A sample of the foreground covariance matrix $\mathbf{F}$ is drawn
using the \texttt{scipy.stats.invwishart} function, using the
$\rm N_\mathrm{pix}\times N_\mathrm{mode}$ matrix of mean-subtracted PCA
amplitudes, the diagonal of which is used for subsequent iterations.

\subsection{The sampling loop}

The outer sampling loop implements the following steps:

\begin{algorithm}[H]
\caption{One Gibbs iteration}
\begin{algorithmic}[1]
\State draw $\boldsymbol{\omega}_s,\boldsymbol{\omega}_f,\boldsymbol{\omega}_d
     \sim \mathcal{N}(0,\mathbf{I})$
\State $\mathbf{b} \gets \texttt{construct\_b}(\mathbf{S},\mathbf{N}^{-1},
     \mathbf{F},w,\overline{\boldsymbol{s}},\overline{\boldsymbol{f}},\mathbf{E},
     \mathbf{d},\boldsymbol{\omega}_s,\boldsymbol{\omega}_f,\boldsymbol{\omega}_d)$
\State $\mathbf{x} \gets \texttt{lgmres}(\mathbf{A},\mathbf{b};
     \mathbf{x}_0=\mathbf{x}^{(r-1)},\,\mathbf{M}^{-1})$
\State unpack $\tilde{\mathbf{s}},\mathbf{f}$ from $\mathbf{x}$
\State $\mathbf{S} \gets \texttt{SCS}(\tilde{\mathbf{s}})$
     \Comment{inverse-gamma per $k$-bin}
\State $\mathbf{F} \gets \texttt{FCS}(\mathbf{f}-\overline{\boldsymbol{f}})$
     \Comment{inverse-Wishart}
\State rebuild $\mathbf{M}^{-1}$ from $(\mathbf{S},\mathbf{F})$
\State save $\mathbf{x},\,P(k),\,\mathbf{S},\,\mathbf{F}$ to disk
\end{algorithmic}
\end{algorithm}

Here, \texttt{SCS} and \texttt{FCS} denote the signal and foreground covariance samplers, respectively. $\textbf{E}$ denotes our set of foreground eigenvectors. Starting points are chosen to be deliberately off-truth: the signal
is initialised to zero, while the foreground amplitudes are
initialised at the true PCA projection perturbed by a 5\% Gaussian
multiplicative jitter, and the initial $P(k)$ is drawn from a
heavily perturbed signal realisation. Each iteration's state
$\mathbf{x}^{(r)}$, sampled $P(k)$, and the diagonals of
$\mathbf{S}$ and $\mathbf{F}$ are saved to disk so that burn-in
and convergence diagnostics can be analysed after the fact. Memory is released explicitly with \texttt{gc.collect()} at the end of every iteration, which we found necessary for chains of $\gtrsim 10^3$ samples at the $128^3$ grid resolution used throughout.

A reference implementation of the Gibbs sampler is available from \url{https://github.com/GeoffMurphy/IM_Gibbs}.


\section{Results}\label{section:Results}

This section presents the results of applying our method to the simulated data, specifically \textsc{Hi} power spectrum and field recovery, both for complete data and channel-flagged data. 

For statistics, namely medians and standard deviations, 2000 samples are drawn for each parameter. We only use the last 1250 samples for our results, with the first 750 being discarded in order to account for any burn-in. While the model was found to converge quickly across all parameters we inspected, for a conservative approach the first 750 samples are always discarded, across all noise levels, in order to account for any parameters that might converge more slowly. This should provide reasonable assurance that convergence has been achieved regardless of noise levels, priors, etc., without having to inspect the trace of each individual parameter (although we have manually inspected traces for a subset of parameters).

\begin{figure}
    \centering
    \includegraphics[width=0.95\columnwidth]{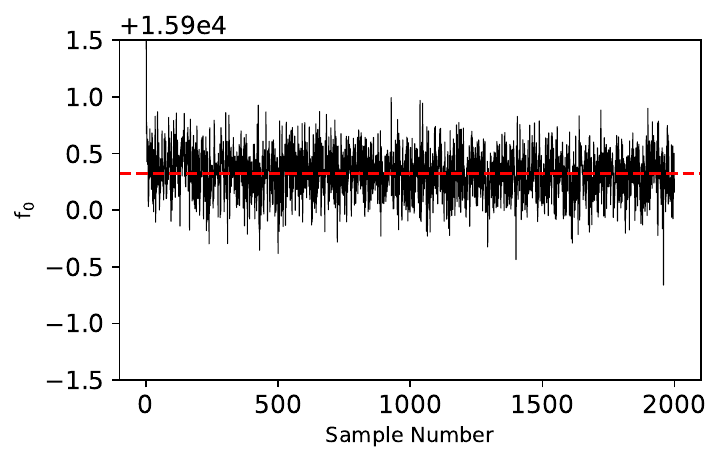}
    \caption{An example of the trace of a foreground mode amplitude to highlight the burn-in period. For all results, we draw a total of 2000 samples, discard the first 750 as burn in, and use the remaining 1250 samples for our statistics.}
    \label{fig:burn_in}
\end{figure}

Fig. \ref{fig:burn_in} shows the trace of a foreground mode amplitude parameter in comparison with its true value. This particular parameter achieves convergence essentially immediately, despite the initial amplitude being well-removed from the true value. The samples can be seen to be moderately correlated, i.e. the effective sample size for this parameter will be less than the raw sample size of 1250 samples. 

We inspected a subset of trace plots to assess that convergence was being achieved across the model. The signal modes and bandpowers were found to be the slowest converging parameters in the model. We also did longer runs of 10,000 samples for the corner plots shown later, and compared the statistics from these to the shorter runs we use for the majority of the results. When comparing the distributions formed from samples $750-2,000$ to those formed from $750-10,000$, we found both cases settled on essentially the same posterior means, with the latter case producing smoother posterior distributions, a result of there being $8,000$ more samples. Also, for the case of $1,250$ retained samples, we found no (unexpected) strong tails or bimodalities in the posteriors, which would have suggested a lack of convergence.

\begin{figure}
    \centering
    \includegraphics[width=0.9\linewidth]{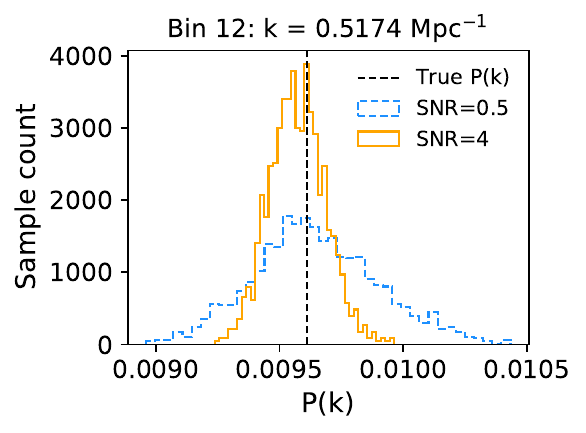}
    \caption{The spread in $P(k)$ samples in the highest $k$-bin for the two noise cases shown here. The vertical black dashed line denotes the true $P(k)$ value. This figure is primarily shown to demonstrate that, while errors are narrow, higher noise cases do induce a wider spread in the $P(k)$ samples.}
    \label{fig:SNR_spread}
\end{figure}

\begin{figure*}
    \centering
    \includegraphics[width=1.95\columnwidth]{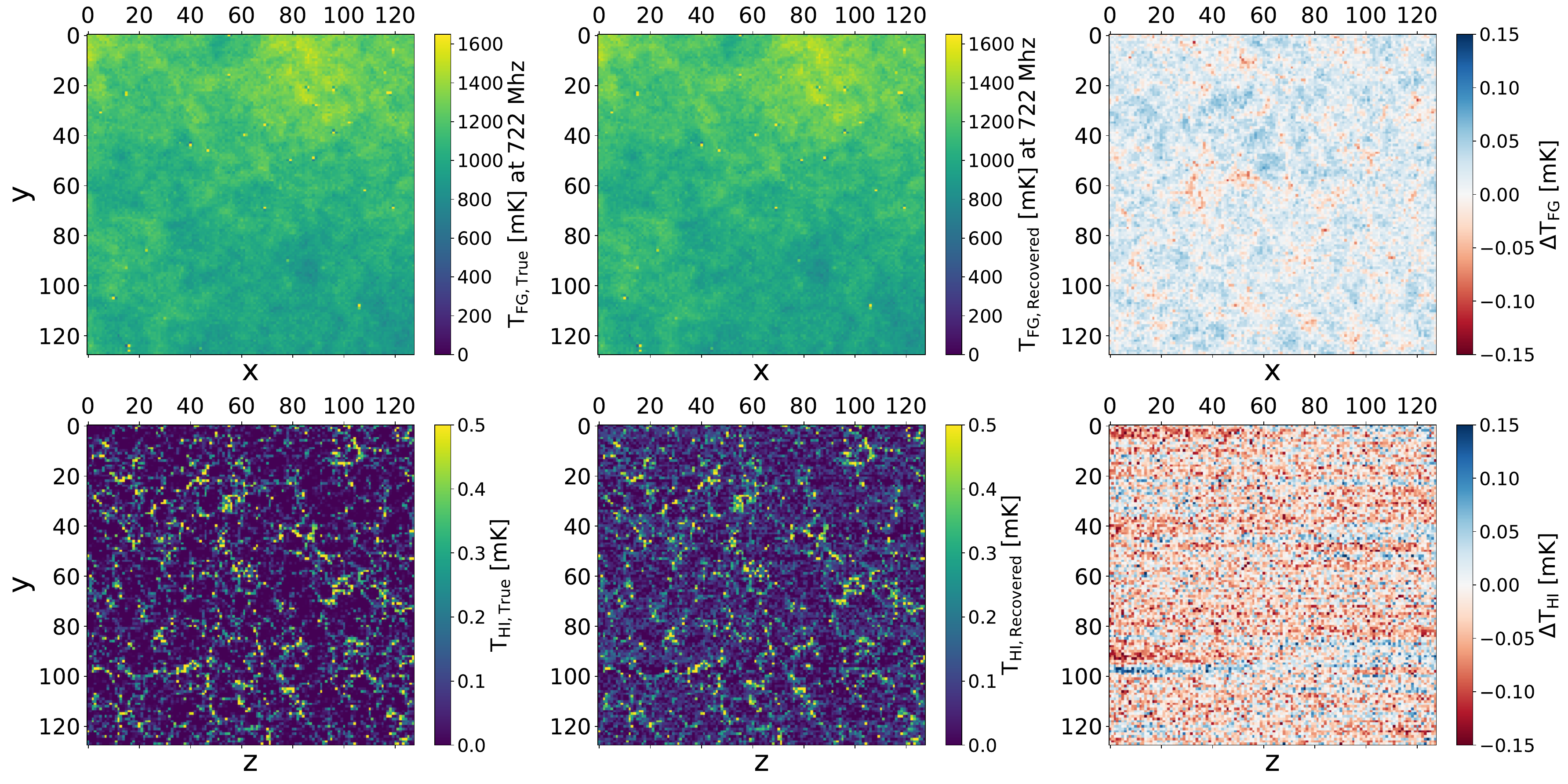}
    \caption{\textit{Upper left:} An example of the foregrounds from the true data cube shown as a slice at a constant frequency of $\nu = 722$~MHz. \textit{Upper middle:} The recovered foregrounds at $\rm{SNR}=4$ (the marginal mean of the foreground component at this frequency). \textit{Upper right:} The foreground residuals, calculated as the difference between the previous two panels. Note the much narrower range of values. \textit{Lower left:} A slice of the (mean-subtracted) true signal field, now in the $y-z$ plane. \textit{Lower middle:} The corresponding slice of the marginal mean of the recovered signal field.  \textit{Lower right:} The residuals of the signal field (the difference of the previous two plots), plotted on the same colour scale as the foreground residuals.}
    \label{fig:cube_rec}
\end{figure*}

\subsection{Power spectrum and 21\:cm field recovery}\label{section:ps_21cm_recovery}

Fig. \ref{fig:recovery_w_fg} shows the recovery of the signal power spectrum achieved by the Gibbs sampler in a scenario with negligible noise. The Gibbs sampler result is shown in orange, with two sample standard deviations being denoted by the error bars. This figure is included to demonstrate the 21\:cm signal recovery in comparison to the foreground power, which is many orders of magnitude larger.

Fig. \ref{fig:snr_comp} shows the spherically-averaged power spectrum recovery as a function of power spectrum signal-to-noise ratio (SNR) in orange, specifically for power spectrum SNRs of $0.5$ and $4$, where the SNR was defined as in Eq.~\ref{eq:noisesnr}, at a particular reference scale of $k = 0.45$~Mpc$^{-1}$. We present this result as a fractional residual for clarity, as the differences between the power spectra are minor. These are the bandpowers drawn from the signal covariance sampler. 

The $\rm SNR = 0.5$ case corresponds to per-voxel noise comparable to the current deepest MeerKLASS L-band deep field ($\rm \sigma_{RMS} \sim 1\,mK$ scale, reported in \citet{Meerklass_2025}), at the noise-dominated end of the $k$-range. The SNR = 4 case is more representative of SKA1-MID single-dish forecasts \citep[e.g.][]{Santos_2015, Bull_2015}, or of a next-generation MeerKLASS campaign combining multiple deep fields.


Fig.~\ref{fig:SNR_spread} shows histograms corresponding to the highest-$k$ bandpower shown in Fig.~\ref{fig:snr_comp} for the same two SNR values. The true power spectrum value is denoted by the vertical black dashed line. This makes it clear that the true value is recovered well in both SNR regimes, but that the statistical uncertainty is larger for the lower SNR, as expected. It is also possible to discern that these distributions are not perfectly Gaussian, e.g. due to their heavier tails and slight skewness. This is also expected; while the amplitude parameters can be marginalised over exactly (given that they are Gaussian), the variance parameters have intrinsic skewness, i.e. inverse-Gamma distributions for the bandpowers and an inverse-Wishart distribution for the foreground covariance modes. These distributions deviate most strongly from their Gaussian limits when the number of modes per bin is small. We see exactly this in our results; Fig. \ref{fig:SNR_spread} shows only slight non-Gaussianity as these are well-populated bins, but in the first two bins of Fig.~\ref{fig:snr_comp} where there are fewer contributing Fourier modes, we see asymmetries in the quoted uncertainties (note the linear scaling in the ordinate axis). These effects are well-characterised in CMB Gibbs sampling applications, for example in \citet{Wandelt_2004} and \citet{Eriksen_2008}, where the $C_\ell$ bandpower distributions also display this skewness.

\begin{figure}
    \centering
    \includegraphics[width=0.65\columnwidth]{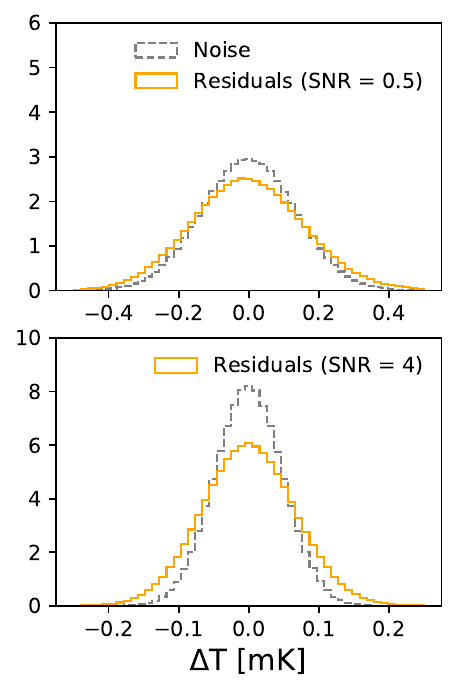}
    \caption{Histograms of the noise, and model cube residuals for SNR = 0.5 (upper plot) and SNR = 4 (lower plot).}
    \label{fig:noise_resids}
\end{figure}


%

Fig. \ref{fig:cube_rec} demonstrates the Gibbs sampler's ability to recover the structure of the foregrounds and signal field at the map level. These results correspond to the {SNR = 4} case shown in Fig.~\ref{fig:snr_comp}. The upper plots compare the true and recovered foregrounds at a single frequency channel, $722\:\rm{MHz}$, while the lower plots compare the 21\:cm signal field recovery along frequency (the $z$ axis). For the former, this is to more easily show the recovered foreground structure. The foreground residuals have correlated structure, but this is small, i.e. at the noise level, as shown in Fig.~\ref{fig:noise_resids}. This is expected, as neither the foreground nor signal components can be recovered perfectly in the presence of noise. The foreground component is biased slightly towards more positive values however.

The signal residual is mostly noise-like, but also shows some structure (streaks) in the frequency direction, and is biased towards slightly negative values. We interpret these features as being caused by the substantial overlap in spectral and spatial shapes between the low-$k$ signal modes and the spectrally-smooth foreground modes -- they cannot be uniquely disentangled from one another, and so are correlated or even degenerate. Hence, a slightly over-estimated foreground temperature is absorbed by a slightly under-estimated signal temperature on large scales, and vice versa.

Fig.~\ref{fig:noise_resids} shows the distribution of the total model (signal plus foreground) residuals compared to the noise, calculated as
\begin{equation}
    \rm \mathbf{T}_{residuals} = \left( \mathbf{T}_{FG}^{True} + \mathbf{T}_{HI}^{True} + \mathbf{T}_{Noise}\right) - \left( \mathbf{T}_{FG}^{Model} + \mathbf{T}_{HI}^{Model}\right),
\end{equation}
where the model values are the marginal means of the Gibbs samples.

In the ideal case, the distribution of the residuals should be similar to the distribution of the noise, which would imply that the sampler is fitting the foregrounds and signal correctly. Overfitting --- where the residuals have a narrower distribution than the noise --- would imply that the model is erroneously absorbing noise into the foreground or signal components, while underfitting (with a wider distribution) would suggest that the model is attributing component power to noise. In the extreme case, this would mean that the noise assumed by the sampler differs significantly from the actual noise.

We show this result for the highest noise case (power spectrum $\rm SNR=0.5$, upper plot), and the lowest noise case ($\rm SNR=4$, lower plot). In both cases, the total residuals are in broad agreement with the noise level, with only slightly broader distributions. This minor difference can mostly be ascribed to imperfect convergence and slight numerical inaccuracies, and is at a level that is essentially negligible.

\begin{figure}
    \centering
    \includegraphics[width=1\columnwidth]{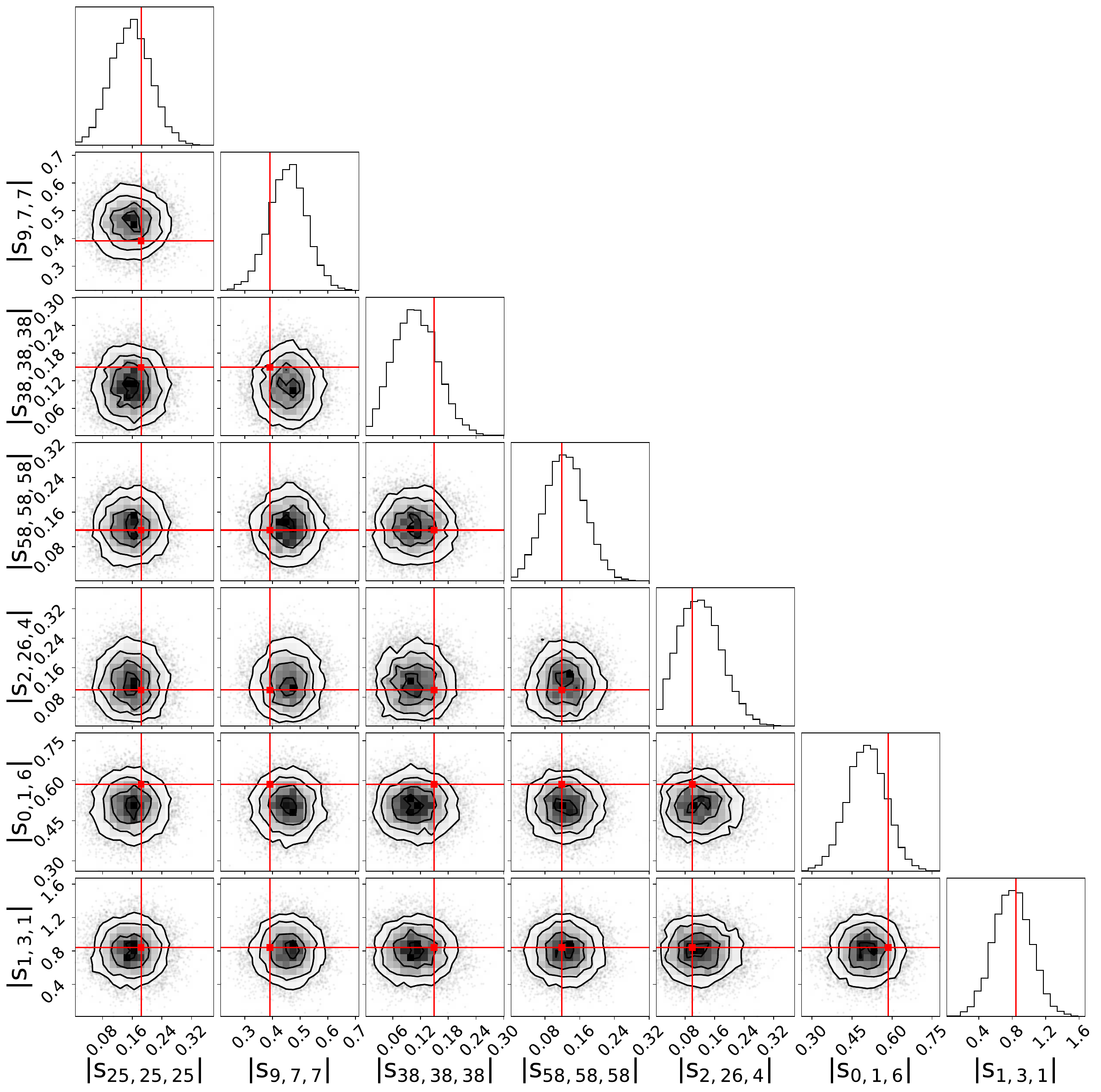}
    \caption{Sample posteriors of seven complex signal Fourier modes, $|\tilde{\mathbf{s}}|$, compared to their corresponding true values. The indices denote Fourier modes in the $65\times128\times128$-shaped \texttt{rfft} cube. This corner plot, and the following ones, are made with $9\:250$ samples (i.e. $10\:000$ samples, with the first $750$ being discarded as burn in), in order to provide a clearer demonstration of the model's convergence. The red lines denote the corresponding true values. }
    \label{fig:s_corner}
\end{figure}

\begin{figure}
    \centering
    \includegraphics[width=1\columnwidth]{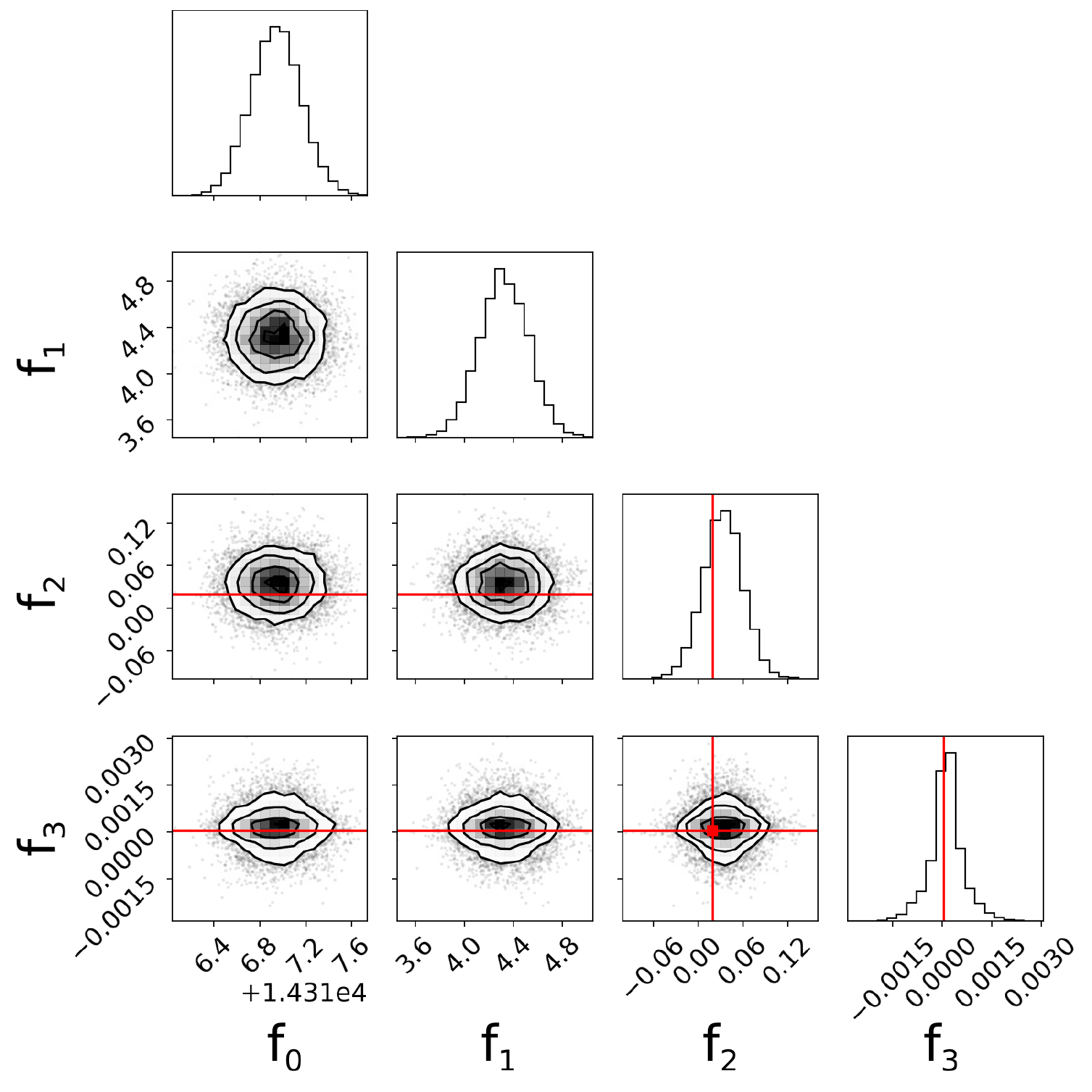}
    \caption{The posteriors of the four foreground mode amplitudes for a particular 1D foreground, compared to their true values. The absence of red lines for certain modes means that the true value was not recovered to high enough accuracy for it to appear on the corner plot.}
    \label{fig:f_amp_corner}
\end{figure}

\begin{figure}
    \centering
    \includegraphics[width=1\columnwidth]{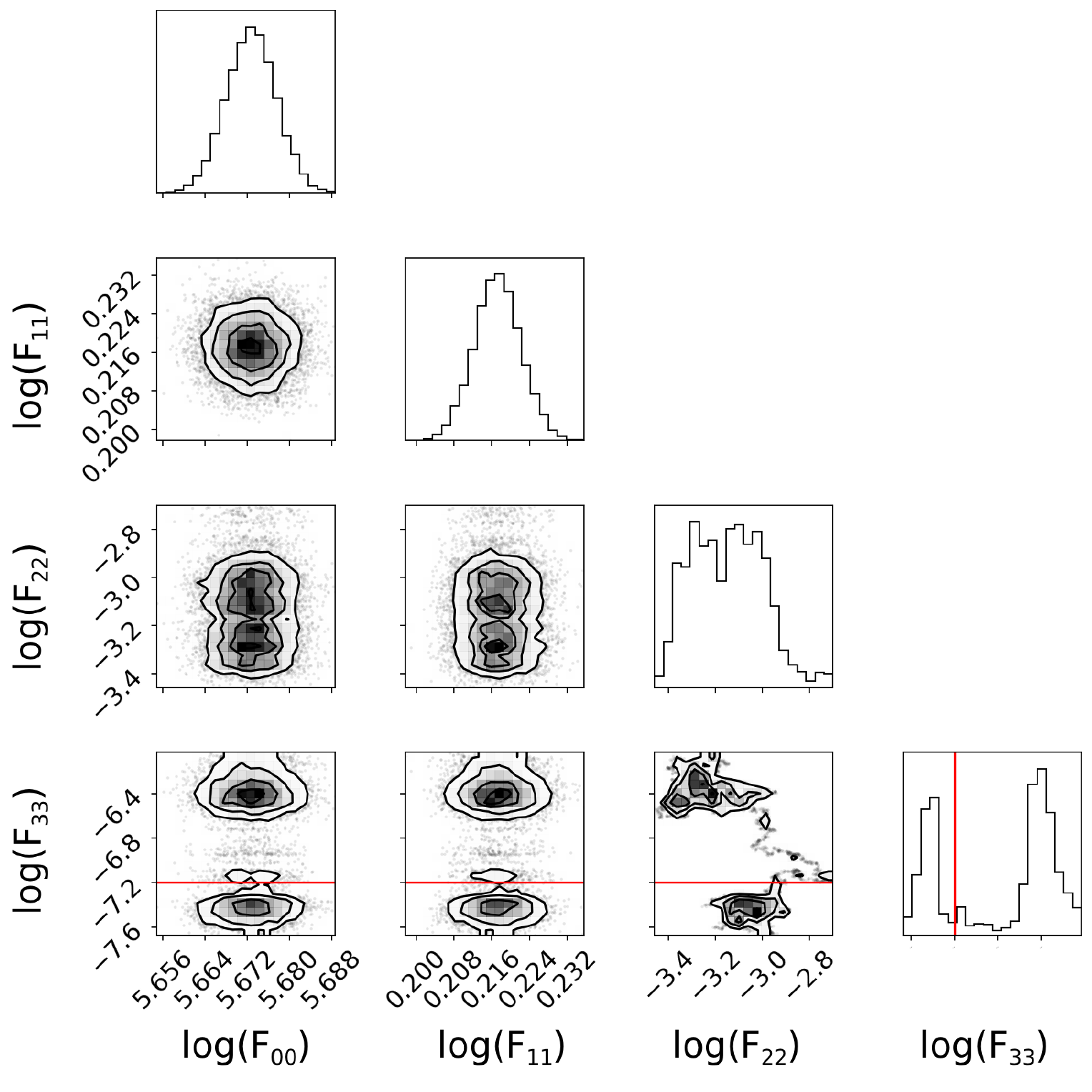}
    \caption{The log posteriors of the four frequency-frequency foreground covariance modes in the model.}
    \label{fig:F_corner}
\end{figure}

Figs.~\ref{fig:s_corner} to \ref{fig:F_corner} show the marginal posteriors for a selection of model parameters, mainly to demonstrate the convergence achieved. This is not exhaustive, particularly for the signal and foreground modes, but should be fairly representative of the $>2$~million model parameters, and serves to demonstrate that the model is fairly well-behaved following burn-in. For these three plots, we ran the sampler for $10,000$ samples and used the same burn-in cut of 750 samples to help identify any convergence issues. For the $2,000$ samples used in all other results, the convergence properties were similar.

\begin{figure*}
    \centering
    \includegraphics[width=1.8\columnwidth]{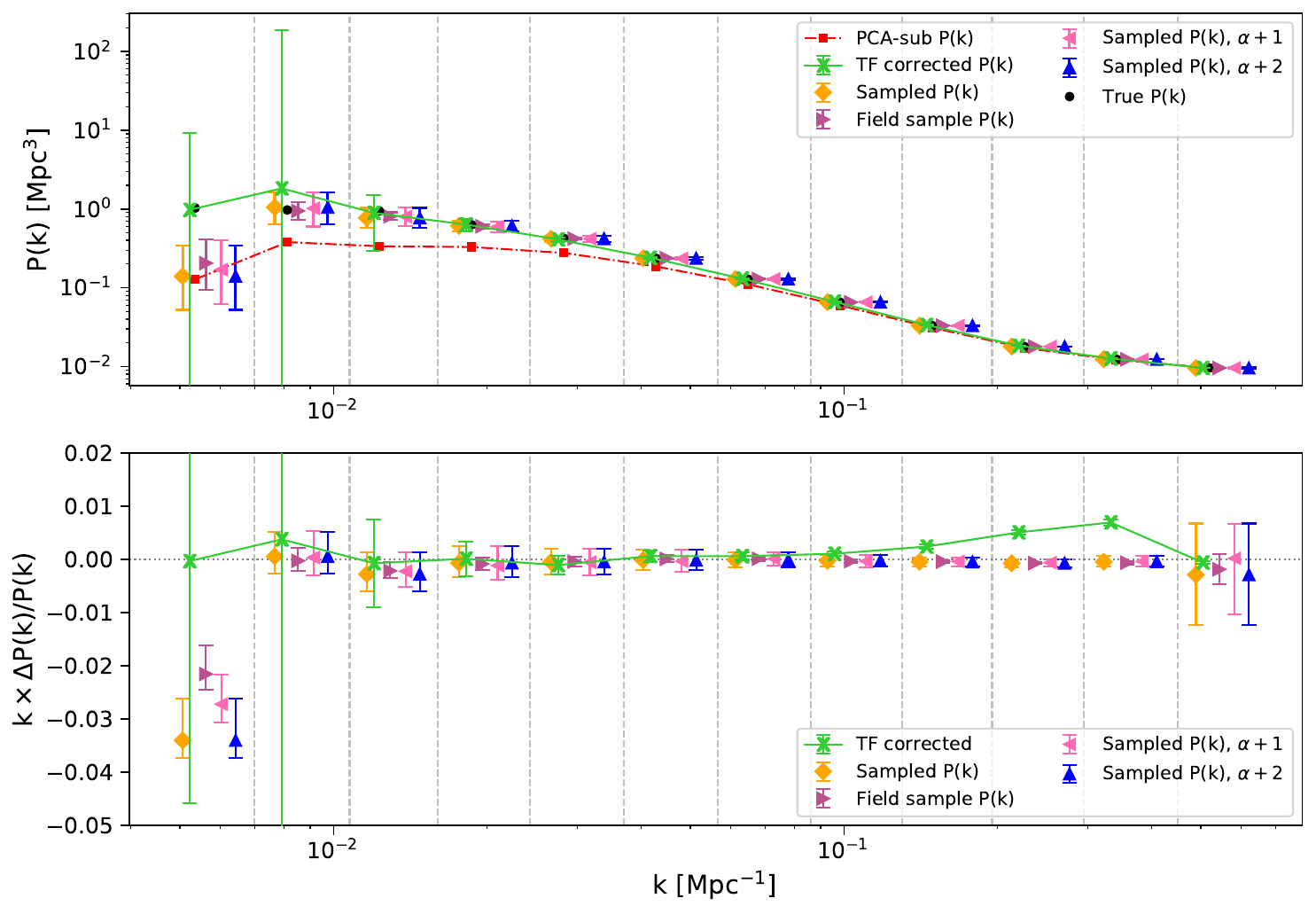}
    \caption{\textit{Top:} Plots of the foreground subtracted power spectrum when blind PCA cleaning is applied in a negligible-noise scenario (red squares), and the subsequent transfer function-corrected power spectrum (green crosses), in comparison to the true power spectrum (black points). Our sample median bandpowers are the orange diamonds, which are from the inverse-Gamma sampler. The dark purple, rightward pointing triangles are instead the bandpowers derived from the model \textsc{Hi} field. Sampled bandpowers using an adjusted prior of $\alpha+1$ (Section \ref{section:GCR_S}) are denoted by the leftward-pointing pink triangles, while $\alpha+2$ is denoted by the upward-pointing blue triangles. Points have been offset for legibility, and vertical grey dashed lines have been added to more easily differentiate between the power spectrum bins. \textit{Bottom:} The fractional residuals of the above results (compared with the true power spectrum), multiplied by $k$ in order to better display the uncertainties over the full range of scales. Errorbars are shown at the 95\% CL.}
    \label{fig:TF_corrected}
\end{figure*}

\begin{figure}
    \centering
    \includegraphics[width=0.99\columnwidth]{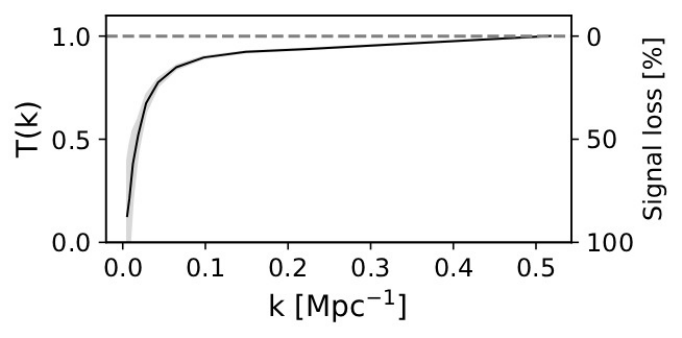}
    \caption{The mean transfer function used to correct the blind foreground cleaned power spectrum in Fig. \ref{fig:TF_corrected}, averaged over 1000 mock realisations of the 21\:cm signal, with the grey region denoting the $2\sigma$ distribution of the realisations. The dashed line denotes $\rm T(k)=1$, or a signal loss of $0\%$.}
    \label{fig:mean_transfer_function}
\end{figure}

Fig.~\ref{fig:s_corner} plots the marginal posteriors of seven signal Fourier mode amplitudes, with their true values shown in red. The 3D mode indices are indicated in the labels, and have been selected at random. The true values of these modes are recovered well in all cases shown here.

Fig.~\ref{fig:f_amp_corner} shows the same for the four foreground mode amplitudes corresponding to a single pixel, and Fig.~\ref{fig:F_corner} the posteriors of all four frequency-frequency foreground covariance parameters. (Recall that the foreground covariance parameters are shared across all pixels.) The picture is more complex for these parameters; the lower order (smoother) foreground mode amplitudes are biased away from the true values, which we expect is due to the partial degeneracy between the signal and foreground modes on large scales discussed above. The mode amplitudes are nevertheless well-measured, and there is no evidence of correlation between the foreground mode amplitudes themselves, consistent with the orthogonality of the chosen basis.

For the foreground covariance parameters in Fig.~\ref{fig:F_corner}, the $F_{33}$ parameter has a bimodal marginal posterior, and shows some evidence of convergence issues, e.g. the faint trail between the modes seen in the $F_{22}-F_{33}$ panel. It is unclear why this bimodality arises, but the modes both appear to be reasonably well explored by the sampler in this case.



\subsection{Comparison to transfer function correction}

We will now compare our results to the commonly-used PCA foreground subtraction method, with a transfer function correction for signal loss. Recall that we described the process of correcting a PCA foreground-subtracted power spectrum in Section~\ref{section:transfer_function}, and that our transfer function is constructed from $1000$ \textsc{FastBox} \textsc{Hi} field realisations.

Fig.~\ref{fig:TF_corrected} shows the results of our sampler, and compares it to both the blind, PCA-cleaned power spectrum, and the transfer function-corrected power spectrum. For reference, the mean transfer function is shown in Fig.~\ref{fig:mean_transfer_function}. For these results, the data being modelled was essentially noise-free (with a power spectrum SNR of $\sim20$).

\begin{figure*}
    \centering
    \includegraphics[width=1.5\columnwidth]{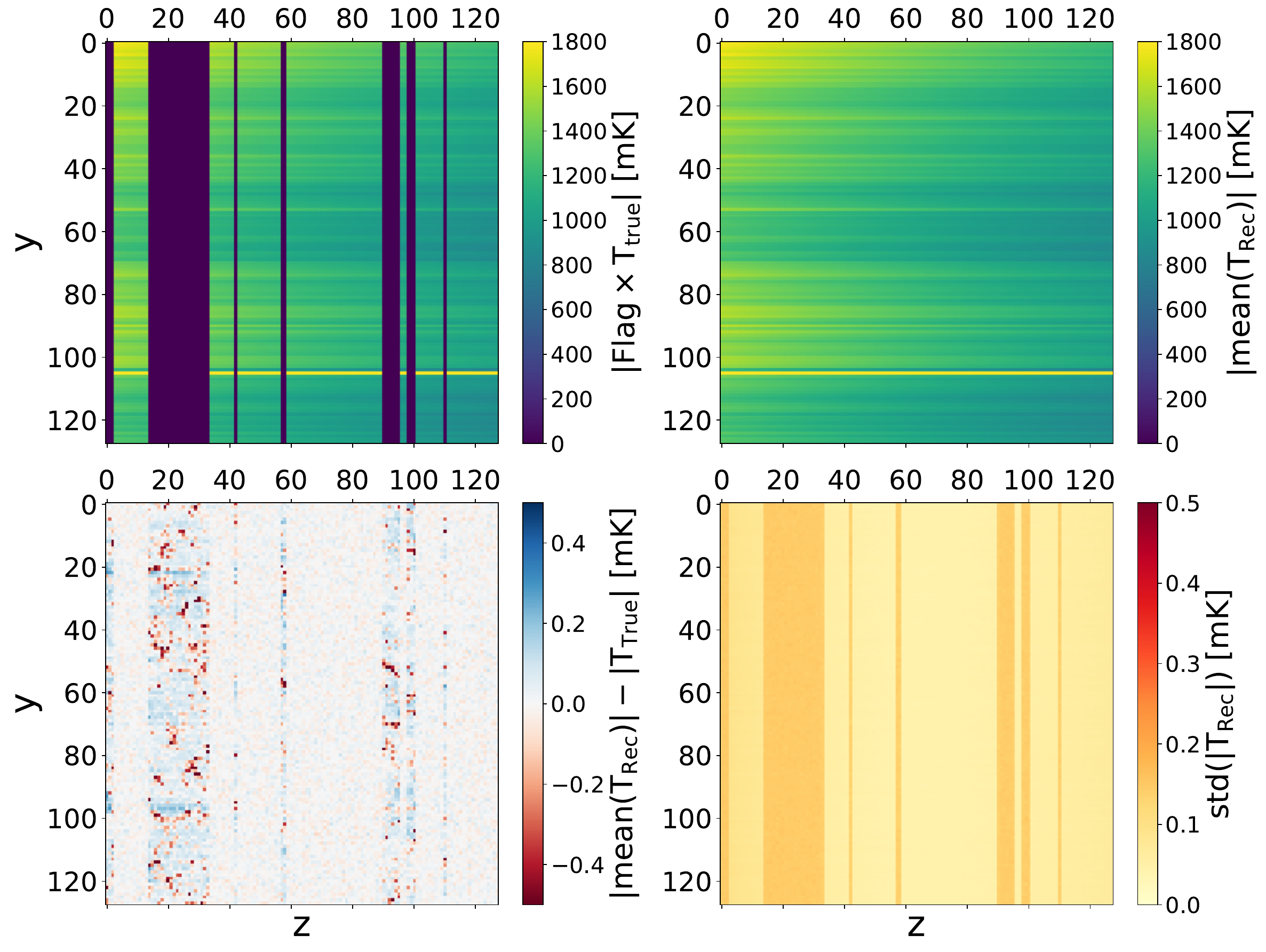}
    \caption{The data cube recovery in the presence of frequency channel flags. \textit{Upper left:} the true data cube (foregrounds, signal, and noise) with flags applied. \textit{Upper right:} The sample mean recovered data cube, where the flagged regions have been in-painted by the model. \textit{Lower left:} The residuals of the model sample mean and the true, unflagged data. \textit{Lower right:} The sample standard deviation of the model data cube.}
    \label{fig:flag_results}
\end{figure*}

\begin{figure*}
    \centering
    \includegraphics[width=1.5\columnwidth]{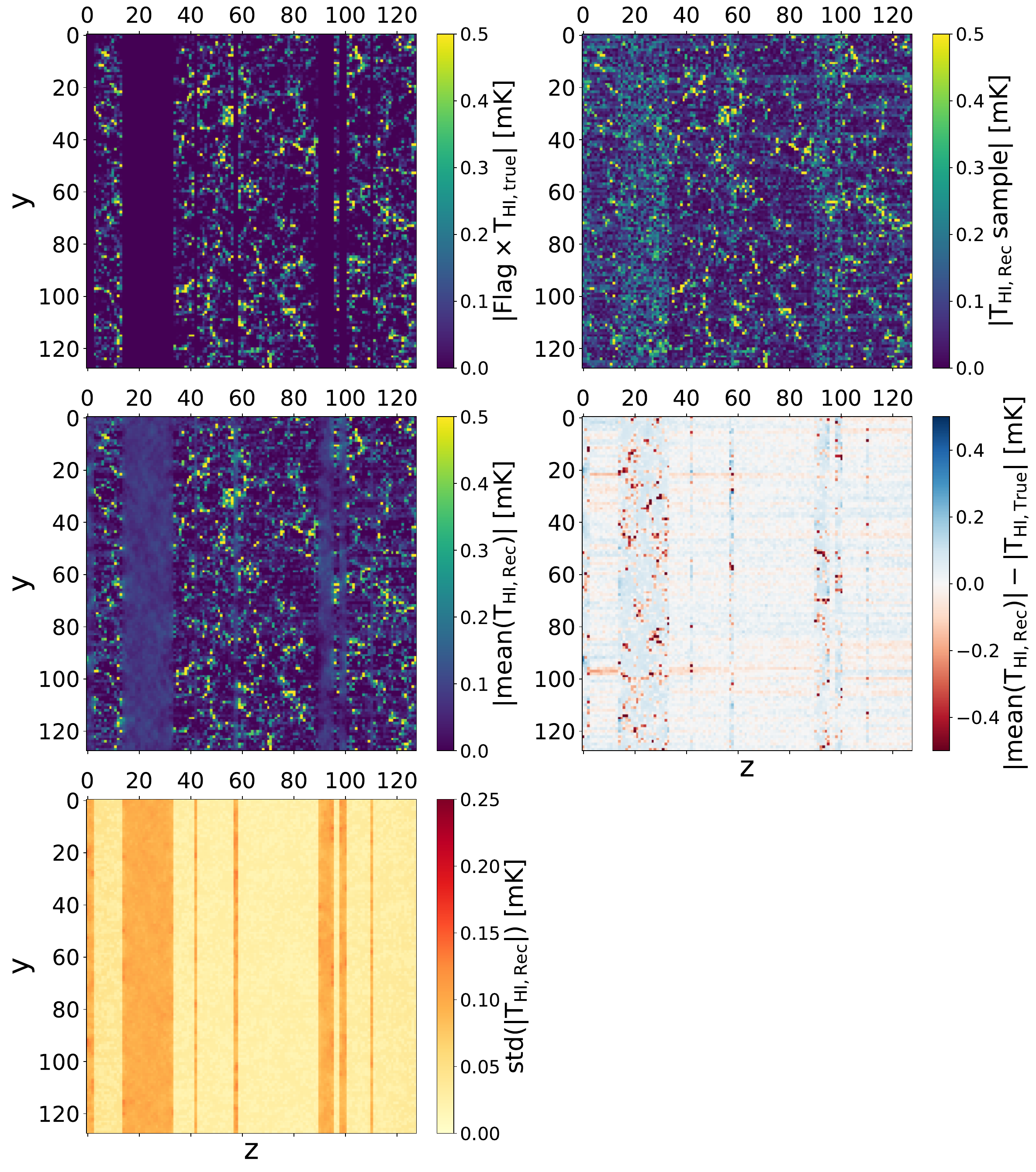}
    \caption{\textit{Top left:} the true 21\:cm field with frequency channel flags applied. \textit{Top right:} A single field realisation from our model, with in-painted flagged channels \textit{Middle left:} the sample mean of our recovered field. \textit{Middle right:} the residuals of the model sample mean and the true, unflagged \textsc{Hi} field. \textit{Bottom:} the sample standard deviation of the model signal field.} 
    \label{fig:flag_signal_results}
\end{figure*}

Our results include four cases. ``Sampled P(k)'' refers to the sampled bandpowers from the inverse Gamma sampler (for the signal covariance). ``Field sample P(k)'' are the bandpowers derived directly from our \textsc{Hi} field samples, i.e. we have $\rm N$ samples of the 3D field, from which we form $\rm N$ `empirical' power spectra, and find the medians and associated uncertainties.
To test prior sensitivity, we also show two cases that adjust the priors on the bandpowers (see Eqs.~\ref{eq:inv_gamma} and \ref{eq:S_prod}). Choosing a prior that increases the inverse Gamma shape parameter $\alpha$ is expected to aid in the separation of foregrounds and signal at large scales, e.g. by reducing the weight of the tails of the inverse Gamma distribution. Our default case is a uniform prior (i.e. $\alpha$ without any correction), but we also test $\alpha+1$ and $\alpha+2$.

From Fig.~\ref{fig:TF_corrected}, it is clear that all of the methods recover the true power spectrum well over the full range of $k$ values, except for the uncorrected PCA method, which suffers from signal loss as expected. The transfer function-corrected result begins to deviate from the true power spectrum at the highest $k$ values, and has larger uncertainties at the lowest $k$ values, but is otherwise robust.

This deviation at high-$k$ is expected. The transfer function applies a single multiplicative correction to each bin and addresses only the amplitude of the signal loss, not the mode-mixing induced by PCA, i.e. residual power leaking between adjacent $k$-bins \citep{2025MNRAS.542L...1C}. This becomes most obvious at high-$k$.

The Gibbs sampler-derived results recover the true power spectrum very well over the entire range of $k$ values, except for in the lowest $k$ bin where a substantial bias is apparent. We anticipate that this bias may reduce for a particularly long run with hundreds of thousands of samples, but a contributor to this is a combination of the small sky coverage ($730\:\rm deg^2$) and the scarcely populated nature of this first bin (18 Fourier modes). This introduces a large sample variance at the scales where the foregrounds are most degenerate with the signal. This degeneracy is seen clearly in a later result where we test power spectrum recovery subject to different numbers of model foreground modes. The primary driver of the bias in this case is the fourth PCA mode suffering a strong degeneracy with the signal at the largest scales.
The starting point of the sampler does not seem to have a strong effect on this bias, as multiple results show the same level of suppression. For example, `Sampled P(k)', the `$\alpha+1$' case, and the `$\alpha+2$' case in Fig.~\ref{fig:TF_corrected} all had distinct starting points, yet show similar biases. Additionally, these PCA modes were formed on the true (simulated) foregrounds, so we can be assured that it is not a model mismatch that is the issue. 

The uncertainties on all four of the Gibbs sampler-derived results are smaller than for the transfer function-corrected result at low $k$, but larger at intermediate and high $k$, reflecting the marginalisation over a more complex model with many parameters. Note that the ``Sampled P(k)'' results have a contribution to the uncertainty from sample variance, as well as the thermal noise, while in principle the ``Field sample P(k)'' does not. The latter does indeed have smaller uncertainties across the range of $k$ values.

We found minimal changes when adjusting the priors on the bandpowers. While the median bandpowers do show some variation, it is not significant enough to suggest that the power spectrum model is prior-sensitive (which is a desirable result).

As briefly mentioned above, recent work in \citet{2025MNRAS.542L...1C} demonstrated that PCA foreground cleaning induces a non‑diagonal window function that mixes 21\:cm signal Fourier modes, and that standard transfer function corrections only fix the amplitude loss rather than the mode‑mixing itself. Because our sampler adopts the same PCA‑derived spectral basis for its foreground model, it should be affected by a similar mode-mixing effect. We expect this effect to naturally manifest as correlations between the signal parameters however, which the Gibbs sampler is capable of exploring efficiently. In other words, the covariance matrix of the power spectrum bandpowers does not need to be separately estimated in our framework; it can be calculated directly from the posterior samples, and is automatically marginalised when we look at the marginal posterior distributions for any of the parameters. We leave a more detailed study of the mode-mixing effect to future work.

\subsection{Frequency channel flagging}

\begin{figure}
    \centering
    \includegraphics[width=0.8\columnwidth]{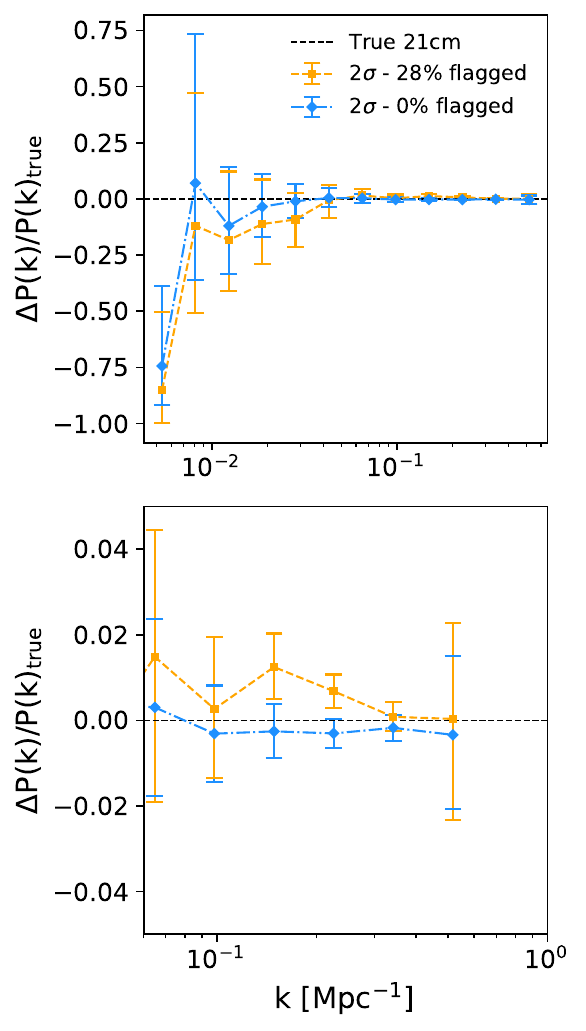}
    \caption{The fractional residuals of the recovered power spectrum in the presence of flagged data compared to complete data. We focus on the high-$k$ modes in the lower plot for visibility. The orange squares denote recovery in the case of $\sim28\%$ of the data volume being flagged, which corresponds to the results of Fig. \ref{fig:flag_signal_results}. The blue diamonds correspond to unflagged data, i.e. the results shown in Fig.~\ref{fig:cube_rec}.}
    \label{fig:pk_flagging}
\end{figure}


We now consider the effect of missing (flagged) data on the recovery of the signal, foregrounds, and power spectrum. The specific flagged frequency channels were not chosen to be entirely representative of that seen in any observations, but were rather chosen to include broad and narrow regions, as well as closely and sparsely-spaced regions. 

Fig.~\ref{fig:flag_results} plots the results for the $\rm{SNR} = 4$ case in the presence of excised data. This figure includes the flagged true data cube (foregrounds, signal, and noise), the recovered model data cube (i.e. the posterior mean averaged over $1250$ samples), the residuals of the recovered model and true (unflagged) data, and the standard deviation of the recovered model cube. Fig.~\ref{fig:flag_signal_results} shows the same analysis, but is limited to the \textsc{Hi} field --- the flagged true field, a single realisation of the recovered field, the posterior mean recovered field, the residuals between the mean recovered and unflagged true fields, and the model standard deviation.


Given the strength of the foregrounds, and the associated ease in modelling this component, it is expected that the recovered data cube (upper right in Fig.~\ref{fig:flag_results}) would in-paint the flagged channels with power consistent with the unflagged ones. The residuals of the data cube do show an increase in discrepancy between the model and true data in the flagged channels, but these are generally small, and are on the order of up to 1\:mK. The sample standard deviation of the model in the flagged channels is also higher than the unflagged channels, which is expected given the lack of data to specifically constrain the model in these regions. Nevertheless, this standard deviation is not greatly increased, being at a level of $\sigma\approx 0.25\: \rm mK$. This indicates that the neighbouring data and smoothness of the model are working to produce a reasonable statistical realisation of the model even in the flagged regions.

More pertinent is the model's recovery of the 21\:cm field, shown in Fig. \ref{fig:flag_signal_results} (top right). In unflagged channels, the model is capable of recovering \textsc{Hi} structure, as before. Furthermore, the model in-paints the flagged channels with statistically-motivated realisations of the 21\:cm field drawn from the power spectrum. It is evident that the in-painted regions are not completely noise-like, with there being correlated structure generated by the sampling. Of course, this structure does not agree with the true \textsc{Hi} field, but it is statistically similar, although not completely indistinguishable, as the flagged regions are still easily identifiable in the figure. We expect this discrepancy to be at least partially caused by the non-Gaussianity of the simulated (true) field, which is not explicitly modelled in the flagged regions. Lastly, the middle-right plot of Fig.~\ref{fig:flag_signal_results} shows the residuals of the mean recovered field, and the bottom plot its standard deviation. The picture is similar to the figure for the total model shown above, with increases in the standard deviation and clear but low-level residual structure in the flagged regions.


As an aside, it can be reasonably expected that the model estimates would worsen as more data is removed. In single dish intensity mapping, however, the same region of sky would be observed by multiple dishes, or by the same dish multiple times. This increase in data redundancy should help to ameliorate the poorer estimates induced by flagging. This will not be the case in frequency channels with persistent RFI however.


Fig.~\ref{fig:pk_flagging} compares the \textsc{Hi} power spectrum recovery in the case of flagged channels (following on from Fig.~\ref{fig:flag_signal_results}), compared to recovery in the absence of flags. We split the $k$-range into two plots for legibility, with the upper plot showing the entire range, and the lower on small scales. 

\begin{figure}
    \centering
    \includegraphics[width=0.9\columnwidth]{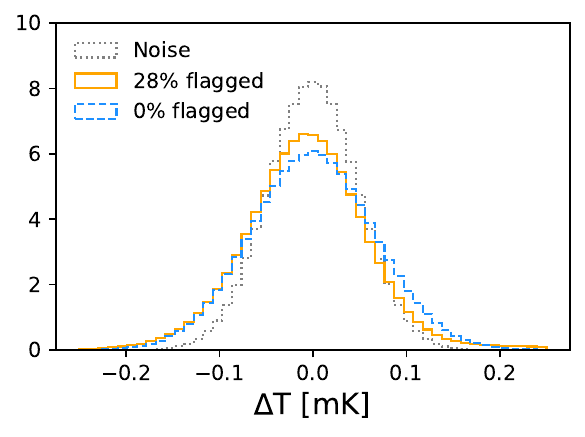}
    \caption{A comparison between the noise and model residuals for the results of Fig. \ref{fig:flag_results}: blue dashed line for the unflagged case, and orange solid for the flagged case, both compared to the noise distribution in the grey dotted line.}
    \label{fig:noise_resids_flag}
\end{figure}

Fig.~\ref{fig:noise_resids_flag} compares the model residuals (note: not the fractional residuals) to the noise for both of these results. 
Recovery of the power spectrum at small scales is quite comparable, suggesting again that even in the presence of missing data, the GCR steps are in-painting with statistically consistent signal realisations. Compared to the noise
distribution, though, both cases do show a slight positive tail, rather than a purely Gaussian distribution. This is expected, given that the simulated \textsc{Hi} field is derived from a by-construction non-Gaussian N-body simulation, and the foregrounds are even more non-Gaussian. In the simulated data, only the noise is Gaussian. The distributions in Fig.~\ref{fig:noise_resids_flag} are most likely to be dominated by the foreground residuals (given the foreground's brightness). Flagging removes large-scale information, worsening constraints and degrading component separation, and results in foreground power leaking into the signal component. When calculating the residuals, this excess power can appear as non-Gaussianities.

On larger scales, recovery of the signal power spectrum remains comparable, apart from the first $k$-bin, where we see a worsening in the underestimation of the power spectrum --- the model median worsens. Given that the first bin only contains 18 modes, and the flagging most significantly removes large-scale information, the impact on large scale recovery is not unexpected. 

We also note that the recovered \textsc{Hi} field in Fig.~\ref{fig:flag_signal_results} shows some artifacts related to partial degeneracies with the foregrounds, most noticeably towards the upper-right corner. These excesses present as horizontal stripes (streaks), and are likely to be related to the slightly poorer estimates of the recovered power spectrum. These seem to be limited to large scales however, are relatively mild, and do not severely affect our estimates in intermediate $k$ bins. We expect that different foreground bases may ameliorate this issue, particularly if correlations between pixels are modelled.



\subsection{Number of foreground modes}

To help understand the impact of our choice of foreground basis on the results, we performed a simple study of changing the number of foreground modes.

\begin{figure}
    \centering
    \includegraphics[width=1\linewidth]{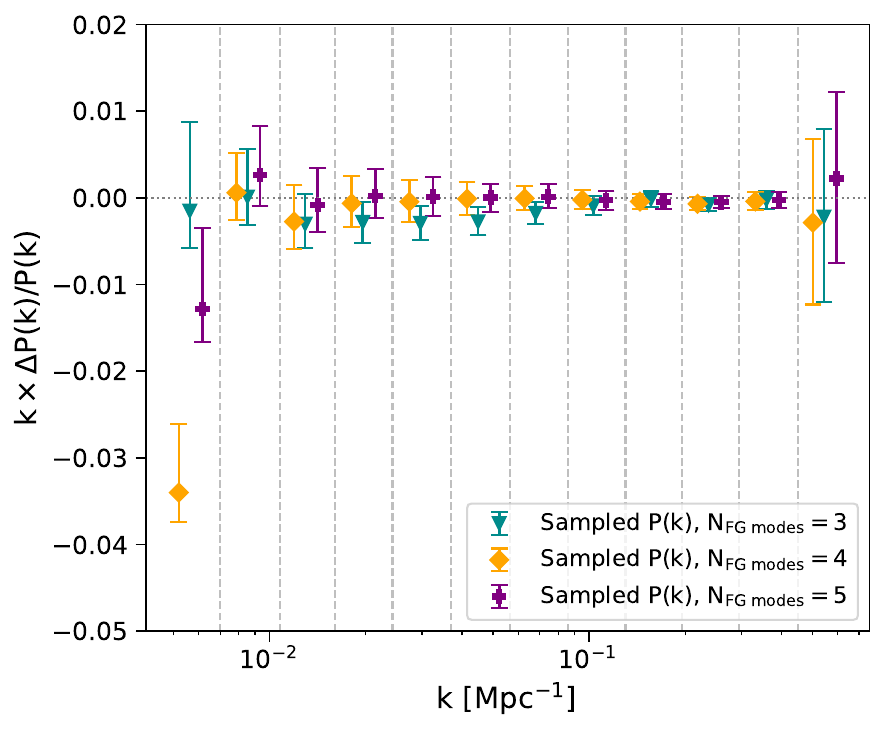}
    \caption{The sampled power spectra for differing numbers of foreground PCA modes: three modes denoted by the downward teal points, four by the orange diamond (the default case), and five by the dark purple plus symbols.}
    \label{fig:n_modes_pk}
\end{figure}

Fig.~\ref{fig:n_modes_pk} plots the sampled power spectra for the cases of $\rm N_{fg} = 3,4,\, and, 5$. Given the eigenvectors shown in Fig. \ref{fig:fg_evecs}, the naïve choice (and the one we use for our results) is four PCA modes. The fourth still contains smooth foreground structure, so is a more `complete' choice compared to $\rm N_{fg} = 3$, while the fifth has noise-like structure that does not seem representative of foregrounds.

We find that both $\rm N_{fg} = 3$ and $5$ provide improved estimates of the lowest $k$ bin. For the case of $\rm N_{fg} = 3$, the implication is that the fourth, unused mode was disproportionately absorbing the \textsc{Hi} signal and thus biasing its power spectrum. However, intermediate power spectrum modes are not recovered as well in this case. For $\rm N_{fg} = 5$, the addition of the noise-like mode also appears to mitigate the large-scale bias somewhat, although the true power spectrum is still not fully recovered. Recovery of other modes are comparable, though, and arguably slightly better in some bins. 


\section{Conclusions}\label{section:Conclusions}
The separation of the \textsc{Hi} signal from bright foregrounds remains an ongoing challenge in observational cosmology, with a variety of methods currently being used in the recovery of the \textsc{Hi} signal. In this work, we propose the application of Gaussian constrained realisations (GCR) and Gibbs sampling to separate foregrounds from the 21\:cm signal at the field level, within a Bayesian statistical framework. Despite the multiple order of magnitude difference between the foreground and \textsc{Hi} signal, in the case of low noise levels, we are able to make an accurate recovery of simulated \textsc{Hi} power spectra on all but the largest scales, as well as recovering field-level structure. 

The Bayesian model that we defined results in $2,195,472$ free parameters. Despite this, we are able to draw samples from the full joint posterior distribution of these parameters at a reasonable speed: $\sim15-30\:\rm s$ per iteration in the case we simulated, depending on the noise and flagging levels, for a data cube of side length $128$. An important factor in achieving this performance are specific modelling choices that ensure that the conditional distributions for large subsets of the parameters are multivariate Gaussian, in which case they can be sampled from directly and efficiently even for a very large number of parameters. This does away with the need to explicitly evaluate a likelihood function with 2 million parameters, something which would be impractical, if not impossible, in any reasonable amount of computational time for most general sampling methods, e.g. MCMC. 

The sampler is capable of recovering the underlying \textsc{Hi} power spectrum accurately, while the sampling approach allows us to form statistical uncertainty estimates on our results directly. For instance, the covariance matrix of the power spectrum bandpowers --- an important quantity for interpreting the power spectrum --- is easily calculated as a natural by-product of the sampler. The statistical model that we defined is found to be quite well-behaved, with convergence being achieved by the sampler in relatively few iterations. 
In comparison with the commonly-used foreground removal approach of subtracting PCA modes and then applying a transfer function correction for signal loss, our method displays similar power spectrum recovery and statistical uncertainties across wavenumber $k$, with only the lowest-$k$ bandpower showing a substantial bias (thought to be due to correlation of these modes with the foregrounds).


The sampler maintains its ability to recover the signal and its power spectrum even when a substantial number of frequency channels are flagged, as is often the case in real data with RFI contamination. When approximately $28\%$ of the channels are flagged, the field level \textsc{Hi} structure can still be recovered in unflagged regions, and both the foregrounds and signal in flagged regions are statistically `in-painted', i.e. the foreground and signal models predict plausible realisations of these components in the flagged region, conditioned on the unflagged data around it. This does away with the need to perform in-painting `by hand', and is instead a natural (and statistically principled) by-product of the GCR steps. This is particularly useful if further harmonic space (e.g. Fourier) analyses are to be applied to the data, as missing data tend to introduce unwanted `ringing' artifacts otherwise. With our method, there is always a complete model prediction that can be used, without any gaps due to flagging.

These results on simulated data are promising, and warrant extension into more realistic scenarios. In future work, we intend to include beam uncertainties, foreground polarisation leakage, and instrumental systematics --- which include broadband RFI, cable reflections, and ground-spill --- among others. These effects would ideally need to be modelled in a manner consistent with the work laid out here, and ideally would be parametrised with linear bases that result in efficient multivariate Gaussian conditional distributions in the Gibbs sampler. If this is not possible, however, then the nature of Gibbs sampling would allow for some parameters to be sampled with GCR, while others could be sampled with more general methods like Markov Chain Monte Carlo or Hamiltonian Monte Carlo, for example. The question will then be whether the mix of samplers and distributions remains efficient enough to be practical.

As we have already discussed, $28\%$ of the frequency channels were excised in our simulated data with flagging. For L-band observations, the MeerKLASS survey with the MeerKAT array considers an observation block to be `good' if less than $50\%$ of its frequency channels are flagged \citep{Meerklass_2025}. Hence, it would be useful to understand the sampler's performance in more difficult flagging scenarios. This also includes flags that vary both with pixel and frequency channel, as well as point source flags (i.e. flags that remove entire sets of pixels). In both cases, the foreground covariance would also need to include spatial information in order for sensible realisations to be drawn in the flagged regions. The foreground covariance model used in this paper did not include spatial (pixel-pixel) correlations however, and so has no way of informing the angular structure of the foreground realisation in flagged regions.

A non-diagonal foreground covariance matrix would be computationally manageable assuming typical mode counts of $\sim10$ (as is used in observational data analysis) and operations would take on the order of a millisecond. A non-diagonal noise covariance would be the next most intensive object to work with, forcing the use of matrix-vector products in our GCR linear equation wherever $\mathbf{N}^{-1}$ is present. Non-diagonality like this would be induced by frequency-frequency correlations from calibration errors, RFI, etc. Even for this work, a full $\sim2\rm M\times2M$ noise covariance matrix would be physically intractable, requiring $\sim32\rm\:TB$ of memory, never mind the memory required for calculations. However, a block-diagonal (or otherwise highly sparse or structured) $\mathbf{N}$ would capture these frequency-frequency correlations while reducing the need for expensive dense matrix products. This implementation would require substantial refactoring compared to what we have currently implemented, but it is still tractable.
We purposely chose to model the \textsc{Hi} signal in Fourier space with a diagonal signal covariance matrix, as is appropriate for a statistically homogeneous and isotropic Gaussian random field. If our (Fourier-space) signal covariance were non-diagonal, even this simple proof-of-concept would again require $\sim32\rm\:TB$ to store $\mathbf{S}$.

Another possible area for improvement is to introduce more realistic beam patterns, with non-trivial frequency dependence and anisotropy for example. This would be an expensive addition, as the forward model for the signal and foregrounds would need to be re-convolved with the beam model at each step. This is relatively inexpensive if the beam is axisymmetric, allowing a Fourier-space convolution to be performed, but for general beams would be a significant bottleneck. For MeerKAT, measured beam patterns do exist \citep{2022AJ....163..135D, 2023AJ....165...78D}, but robust uncertainty estimates and suitable basis expansions would be needed to use these as the foundation of a beam forward model. Typical beam patterns are rather complex however, and need carefully-chosen basis functions. See \citet{2024RASTI...3..400W, 2025RASTI...4...42W} for examples of implementing a beam and foreground Gibbs sampler in the context of interferometric measurements.

Holistic Gibbs samplers that ingest time-ordered data and infer signal, foreground, and instrumental model components in a single joint inference scheme have been successfully applied to CMB data \citep[e.g.][]{2023A&A...679A.143W}. Given the additional complexity of 21 cm data, we advocate for a two-step process instead, where time-ordered data can be processed into realisations of calibrated sky maps using a statistical Gibbs sampling approach such as the one recently presented by \citet{2026RASTI...5ag024Z}, which are then passed into the map-level analysis framework we have presented here. The posterior uncertainties from the first step would need to be propagated, e.g. into an effective noise variance map, but a variety of low-level instrumental effects such as beam model uncertainties, polarisation leakage, and correlated noise could be marginalised before handing the `cleaned' map data to the next step, significantly reducing the total computational complexity. This would not permit the {\it full} joint posterior of all of the signal, foreground, and systematics components to be recovered, however.

Our more immediate aim is to model the signal in all elements of a large area survey akin to MeerKLASS, something that would not be feasible with most other sampling methods. For MeerKAT's $64$ dishes, it should be fairly straightforward to extend our framework to include the per-antenna data for all of the antennas simultaneously (i.e. without simply averaging over antennas). Since each dish is observing the same sky region, they would all share the same signal and foreground model, with only the differences in flagging and systematics needing to be taken into account on a per-antenna basis. 

A single sampling iteration in this work takes around 20 seconds. For 64 dishes, substantially more memory and computing time would be needed, as some of the linear systems used by the sampler would be much larger. It would be possible to take advantage of parallel processing to reduce the wall-clock time required by the sampler, and split the data structures across multiple compute nodes, and so we do not anticipate any fundamental barriers to increasing the dimensionality of the model even further.

\section*{Acknowledgements}
GGM acknowledges support from the South African National Research Foundation (Grant No. SPDF240806256553).
MGS acknowledges support from the South African National Research Foundation (Grant No. 84156).
This result is part of a project that has received funding from the European Research Council (ERC) under the European Union's Horizon 2020 research and innovation programme (Grant agreement No. 948764; PB, ZZ).
SC acknowledges support from the UKRI Stephen Hawking Fellowship (grant reference EP/U536751/1) and a UK Research and Innovation Future Leaders Fellowship grant [MR/V026437/1].

We acknowledge the use of the ilifu cloud computing facility – www.ilifu.ac.za, a partnership between the University of Cape Town, the University of the Western Cape, Stellenbosch University, Sol Plaatje University and the Cape Peninsula University of Technology. The ilifu facility is supported by contributions from the Inter-University Institute for Data Intensive Astronomy (IDIA – a partnership between the University of Cape Town, the University of Pretoria and the University of the Western Cape), the Computational Biology division at UCT and the Data Intensive Research Initiative of South Africa (DIRISA).

We acknowledge the use of the following software: \textsc{numpy} \citep{Harris_2020}, \textsc{matplotlib} \citep{Hunter_2007}, \textsc{scipy} \citep{Virtanen_2020}, and \textsc{corner} \citep{Foreman-Mackey_2016}.
\section*{Data Availability}

Example analysis scripts used in this work are available from \url{https://github.com/GeoffMurphy/IM_Gibbs}. Data outputs presented in this paper are available upon reasonable request from the corresponding author.

\balance



\bibliographystyle{mnras}
\bibliography{joint_bayes} 







\bsp	
\label{lastpage}
\end{document}